\documentclass{article}

\usepackage{arxiv}
\usepackage{amsmath}
\usepackage{graphicx}
\usepackage{epstopdf}
\usepackage[utf8]{inputenc} 
\usepackage[T1]{fontenc}    
\usepackage{hyperref}       
\usepackage{url}            
\usepackage{booktabs}       
\usepackage{amsfonts}       
\usepackage{nicefrac}       
\usepackage{microtype}      
\usepackage{lipsum}
\usepackage{algorithm}
\usepackage{algorithmic}
\usepackage{booktabs}
\usepackage{lscape}
\usepackage{caption}
\usepackage{subcaption}
\usepackage{multirow}

\newtheorem{defi}{Definition}[section]
\newtheorem{thm}{Theorem}[section]

\newcommand{\R}{\ensuremath{\mathbb{R}}}

\newcommand{\goto}{\ensuremath{\rightarrow}}

\newenvironment{Assumptions}
{
\setcounter{enumi}{0}

\begin{enumerate}}
{\end{enumerate} }

\title{A Fuzzy Edge Detector Driven Telegraph Total Variation Model For Image Despeckling}

\author{
 Sudeb Majee \\
  School of Basic Sciences\\ 
  Indian Institute of Technology Mandi\\
  PIN 175005, INDIA\\
  \texttt{sudebmajee@gmail.com} \\
    \And
     Subit K. Jain \\
  School of Basic Sciences\\ 
  Indian Institute of Technology Mandi\\
  PIN 175005, INDIA\\
  \texttt{jain.subit@gmail.com} \\
   \and
  Rajendra K. Ray \\
  School of Basic Sciences\\ 
  Indian Institute of Technology Mandi\\
  PIN 175005, INDIA\\
  \texttt{rajendra@iitmandi.ac.in} \\
    \And
   Ananta K. Majee \\
   Department of Mathematics\\
   Indian Institute of Technology Delhi\\
   PIN 110016, INDIA \\
  \texttt{majee@maths.iitd.ac.in} \\ 
}

\begin{document}
\maketitle

\begin{abstract}
Speckle noise suppression is a challenging and crucial pre-processing stage for higher level image analysis. In this work, a new attempt has been made using telegraph total variation equation and fuzzy set theory for speckle noise suppression. The intuitionistic fuzzy divergence (IFD) function has been used to distinguish between edges and noise. To the best of the author's knowledge, most of the studies on multiplicative speckle noise removal process focus on only diffusion based filters, and little attention has been paid to the study of fuzzy set theory. The proposed approach enjoy the benefits of both telegraph total variation equation and fuzzy edge detector, which is not only robust to noise but also preserves image structural details. Moreover, we establish the existence and uniqueness of a weak solution of the regularized version of the proposed model using Schauder fixed point theorem. With the proposed model, despeckling is carried out on natural and Synthetic Aperture Radar (SAR) images. The experimental results of the proposed model are reported, which found better in terms of noise suppression and detail/edge preservation, with respect to the existing approaches.
\end{abstract}

\section{Introduction}
\label{intro}
Speckle noise distorts edge/texture and subtle details of digital images (i.e., synthetic aperture radar images, ultrasound images, and laser images), which may contain meaningful statistics \cite{burckhardt1978speckle}. Its appearance in images reduces the utility and detectability of objects in the image. Therefore, it is necessary to develop and implement a novel image despeckling approach that can enhance the visual quality before commencing high-level image analysis.

In the last few decades, a rising figure of studies describes the statistical and the fundamentals properties of the multiplicative speckle noise, usually for the synthetic aperture radar (SAR) and medical ultrasound images \cite{jain2019non}. 
Among the popular state-of-the-art despeckling approaches, anisotropic diffusion based partial differential equation (PDE) methods \cite{jain2018nonlinear,jin2000adaptive,shan2019multiplicative,yu2002speckle,zhou2015doubly} and variational based methods \cite{aubert2008variational,dong2013convex,jin2011variational, liu2013nondivergence,rudin2003multiplicative} are widely used to formulate the speckle noise removal strategies. The first variational approach to deal with multiplicative noise removal problem is given by Rudin et al. \cite{rudin2003multiplicative}, which is known as RLO model. This filter provides improved results by protecting some important details from over-smoothing. The significant deficiency of this model is that it does not use the notion of gamma noise therefore not guaranteed to present better results. Aubert and Aujol \cite{aubert2008variational} introduced a non-convex model(AA model) by utilizing maximum a posteriori (MAP) estimator for the multiplicative Gamma noise. 
As far as we know that, most of the researchers showing their interest in parabolic type PDEs for image despeckling problem. However, the PDEs of hyperbolic type \cite{ratner2007image} could upgrade the visual appearance of the located edges and improve their quality better than parabolic PDEs.
The PDE based diffusion models are efficient in image restoration. But under high noise levels uncertainty emerges in the classification of the clean and affected pixels, which give an edge map with multiple false positive edge pixels. To overcome this issue of uncertainty and to obtain a meaningful decision, another emerging technique namely fuzzy set theory is used for noise removal problems \cite{babu2016adaptive,binaee2014ultrasound,hua2009speckle,nadeem2019fuzzy}. Fuzzy logic based approach is basically a two-stage architecture; at first classify each pixel in an image into three different classes such as `edge', `noise' and `homogenous', and then performs fuzzy filtering by using these detected edges and their impact on the neighboring pixel values. Both steps are based on the fuzzy set theory which makes use of membership functions \cite{klir1995fuzzy}.
In \cite{aja2001fuzzy,prasath2014image,song2003fuzzy}, some hybrid fuzzy anisotropic diffusion methods are studied. In these hybrid techniques, fuzzy logic based diffusion coefficients are used to control the whole diffusion process, instead of taking image gradient dependent diffusion coefficient. 
A major problem with any image restoration algorithm based on regular anisotropic diffusion is lack of the description of the statistical properties to the present degradation. The variational technique is more suitable for this issue. It is surprising to note that there is no fuzzy edge detector \ref{sec:IFS} based total variation model \ref{subsec:CMMN} with telegraph diffusion \ref{subsec:TDM} framework for speckle noise suppression with effective edge preservation, yet.

Hence, to prove the efficacy of the telegraph equation for speckle noise reduction, the present work provides a new approach by viewing the image as an elastic sheet \cite{ratner2007image} in the total variation framework \cite{dong2013convex}. Moreover, we use a fuzzy edge detector function \cite{chaira2008new} which is very efficient for image edge detection.
There are two key advantages of this proposed approach. First, we use telegraph equation \cite{ratner2007image}, derive from the total variation framework \ref{sec:our model}, which can provide sharp and true edges better than other non-telegraph based total variation algorithms during the noise removal process. Second, a fuzzy template based edge detector function is incorporated into the telegraph total variation framework due to its effective edge/noise separation ability. Furthermore, we study the existence and uniqueness of a weak solution of the regularized version of the proposed model using Schauder fixed point theorem on an appropriate function space. Finally, the present model applied to some standard natural test images along with different real SAR images corrupted by speckle noise. The latter is typically deliberated, as the presence of speckle noise is an indispensable characteristic of SAR image. This type of noise makes target extraction and analysis of objects more stiff and erratic. Therefore, to enhance the observation of SAR images without violating actuality and textures, development of new speckle noise suppression algorithms play an important role. The image quality of despeckled images utilizing the proposed model has been compared with some existing PDE based models.

The rest of the paper is organized as follows. Section \ref{sec:Material and Methods} introduces the details of the telegraph-diffusion filter,  the convex multiplicative model and the concept of fuzzy based edge detection. Section \ref{sec:Proposed Model} describes the proposed fuzzy edge detector induced telegraph total variation method for removal of speckle noise. In the section, \ref{sec:analysis} we study the existence and uniqueness of a weak solution of the regularized version of the proposed model. Section \ref{sec:numerical} describes the numerical implementation of the proposed approach. The simulated despeckling results obtained by the proposed approach are compared with other discussed diffusion methods in Section \ref{sec:Results}. We conclude the paper in Section \ref{sec:Conclusion} with an outlook on future work.

\section{Material and Methods}
\label{sec:Material and Methods}
\subsection{Telegraph-Diffusion Model}
\label{subsec:TDM}
 In the existing literature, the first hyperbolic model for image denoising is telegraph-diffusion model \cite{ratner2007image}, where consider the image as an elastic sheet, which interpolates between the parabolic PDE and hyperbolic PDE. The telegraph-diffusion model takes the form,
\begin{align*}
u_{tt}+\gamma u_t  = &\text{div}(c_1(|\nabla u|)\nabla u),  \hspace{1.0cm} \text{in} \hspace{0.2cm} \Omega_T:=\Omega \times (0,T),  \\
&\frac{\partial u}{\partial n}=0,              \hspace{2.3cm} \text{in}
\hspace{0.2cm} \partial\Omega_T:= \partial \Omega \times (0,T),\\
u(x,0)=&u_0(x), \hspace{0.2cm} u_t(x,0)=0,	\hspace{0.5cm} \text{in} \hspace{0.2cm}\Omega,
\end{align*}
where $\Omega$ is the domain of original image $u$ and the observed noise image $u_0$, $\text{div}$ and $\nabla$ represents the divergence and gradient operator respectively. $c_1(s)=1/(1+({s^2}/{k^2}))$ is an edge-controlled diffusion function which preserves the important features and smoothens the unwanted signals, and $\gamma$ is the damping parameter.
%
%
Even though the TDE model can effectively preserve the sharp edges during the smoothing of noisy data, it is unstable and failed to produce satisfactory smoothing in the presence of large noise level. To overcome this issue, several non-linear telegraph diffusion models \cite{cao2010class,jain2016edge,sun2016class} have been proposed by many researchers. To the best of our knowledge, in spite of their imposing applications in the area of additive noise elimination, hyperbolic PDEs based approaches have not yet been studied for multiplicative speckle noise removal.

\subsection{A Convex Model for Multiplicative Noise}
\label{subsec:CMMN}
In general, the diffusion processes in image processing have its origin in the variational calculus \cite{aubert2006mathematical}. Typically, these variational approaches can be defined as a combination of fidelity and regularization term. The first variational calculus based approach for suppression of multiplicative noise is given by Rudin, Lions, and Osher and termed as RLO model \cite{rudin2003multiplicative}.

Aubert and Aujol \cite{aubert2008variational} introduced a new functional model (AA model) with the fidelity term which is strictly convex for $I \in (0, 2I_0)$.

To overcome the non-convexity and locally optimal solution issues of the RLO model and AA model respectively, several authors have used a convex function with different data fidelity terms \cite{jin2011variational, liu2013nondivergence}. Recently, Dong et al. \cite{dong2013convex} introduce the following optimization problem for multiplicative speckle noise elimination:
\begin{equation}\label{eq:energy_direct}
 I=\displaystyle \min_{I \in \text{BV}(\Omega)} \lbrace J(I) + \lambda H(I,I_0)\rbrace,  \nonumber
\end{equation}
where $J(I)$ represents the total variation of $I$ and $H(I, I_0)$ is a fidelity term given by
\begin{equation*}
H(I,I_0)=\int_{\Omega}\left(I+I_0 \log\left(\frac{1}{I}\right)\right)dx.
\end{equation*}
In \cite{dong2013convex}, the authors have established the existence and uniqueness of the minimizer for the following energy functional for removal of speckle noise,
\begin{equation}\label{eq:Dong_energy}
\min_{I\in \text{BV} \left(\Omega\right)}\left\lbrace \int_{\Omega} \alpha(x)|\nabla I|dx+\lambda \int_{\Omega}\left( I+I_0 \log\frac{1}{I}\right) dx\right\rbrace \nonumber
\end{equation} 
where the value of gray level indicator was adopted as,
\begin{equation}\label{eq:gray_indicator}
\alpha(x) = \left( 1-\frac{1}{1+k|G_\xi \ast I_0|^2}\right) \frac{1+kM^2}{kM^2}, \nonumber
\end{equation}
with $M=\underset{x \in \Omega}{\text{sup}}(G_\xi \ast I_0)(x)$ where, $\xi>0$, $k>0$, and ``$\ast$" represents the convolution operator and $G_\xi$ is the two dimensional Gaussian kernel.
\subsection{Intuitionistic Fuzzy Sets}
\label{sec:IFS}
The existing speckle noise filtering approaches fail to preserve the significant information: namely to capture the edge information from noise, thereby suppressing the edges or enhancing the noise particle assuming by edges. Therefore, to preserve the image details along with smoothing, the theory of fuzzy set has gained much popularity in recent times \cite{becerikli2005new,chaira2003segmentation,ho1995fedge}. Atanassov \cite{atanassov2003intuitionistic} proposed the concept of intuitionistic fuzzy set in which the author combined the degree of non-membership with some hesitation degree. Szmidt and Kacpryzk \cite{szmidt2000distances} introduced some new distance measures between intuitionistic fuzzy sets with the generalization of the Hamming and the Euclidean distance. Next, using intuitionistic fuzzy sets, Chaira et al. \cite{chaira2008new} proposed a new measure, called intuitionistic fuzzy divergence (IFD), and its application to edge detection.
A subset or fuzzy set $P$ in a finite universal set $Y = \left\lbrace y_1, y_2, . . ., y_n \right\rbrace $ may be defined as
\begin{equation*}
P= \left\lbrace (y, \mu_P(y))| y \in Y \right\rbrace,
\end{equation*}
where the characteristic or membership function $\mu_P(y)$ represents the measure of belonging-ness of an element $y$ in the finite set $Y$. This degree of membership defined on P, assumes a characteristic value between $0$ and $1$ i.e. $\mu_P(y) \in [0, 1]$. An intuitionistic fuzzy set, proposed by Attanassov, may be mathematically given as
\begin{equation*}
P= \left\lbrace (y, \mu_P(y), \nu_P(y))| y \in Y \right\rbrace,
\end{equation*}
where $\mu_P(y):Y \rightarrow [0, 1]$ and $\nu_P(y):Y \rightarrow [0, 1]$ represent degree of membership and non-membership, respectively, with the necessary condition
\begin{equation*}
0 \leq \mu_P(y) + \nu_P(y) \leq 1.
\end{equation*}
Further, a third parameter $\pi_P(y)$ is considered which is known as the hesitation degree. Recently \cite{chaira2008new}, with the consideration of the membership degree, the non-membership degree, and the hesitation degree, intuitionistic fuzzy divergence (IFD) is defined as 
\begin{align*}
Div\_measure(j,k)=\max_{N}[\min_{r}(\text{Div}(P,Q))],
\end{align*}
where $P$ and $Q$ represents two intuitionistic fuzzy sets, where $N$ is the number of templates and $r$ is the number of elements in the template. The fuzzy divergence between the elements $p_{j,k}$ and $q_{j,k}$ of image set P and template set Q, Div(P, Q), is calculated as,
\begin{equation*}
\begin{aligned}
F(I)&=\text{Div}(p_{j,k},q_{j,k}) \\
& = \big\{2-(1-\mu_P(p_{j,k})+\mu_Q(q_{j,k}))e^{\mu_P(p_{j,k})-\mu_Q(q_{j,k})} \\
      & -(1-\mu_Q(q_{j,k})+\mu_P(p_{j,k}))e^{\mu_Q(q_{j,k})-\mu_P(p_{j,k})}\big\}.
\end{aligned}
\end{equation*}

\section{A New Telegraph Total Variation Multiplicative Model for Speckle Noise Removal}
\label{sec:Proposed Model}
\subsection{The Proposed Model}
\label{sec:our model}
Following the idea of  \cite{dong2013convex} consider the energy minimization problem:
\begin{equation*}\label{eq:new_energy}
I^{*}=\underset{I\in \Omega}{\text{argmin}}    \left\lbrace \int_{\Omega} TV(I) +\lambda \int_{\Omega}\left( I+I_0 \log\frac{1}{I}\right) dx\right\rbrace, 
\end{equation*}  
where $\Omega=\left\lbrace I>0, I \in BV(\Omega) \right\rbrace$ and $TV(I)$ is finite total variation in $I$. The fidelity term given in the above minimization problem is strictly convex for all $I$. Let us now consider the following Euler-Lagrange equation for some total variation problem, 
\begin{align*}
 G(\nabla I, {\nabla}^2 I)-\lambda \left( 1-\frac{I_0}{I}\right)&=0,\,\, \text{in} \,\, \Omega,  \\
\frac{\partial I}{\partial n} &=0,\,\, \text{in} \,\, \partial \Omega, \\
\end{align*}
where $\nabla I$ and ${\nabla}^2 I$ represents the gradient and Hessian matrix of $I$. Note that the obtained edge information highly depends on the magnitude of the gradient, which produces broken and discontinued edges. As image edges and boundaries have fuzziness, which is caused by multiplicative speckle noise, we propose the following fuzzy edge detector driven total variation model:
\begin{equation*}
\min \int_{\Omega} \theta(I)|\nabla I|dx,
\end{equation*}  
where the fuzzy edge indicator functions $\theta(I)= 1-F(I)$ controls the amount of smoothing at different regions by providing pixel-wise edge characterization using the IFD function. Hence, with the choice of positive valued function $\theta(I)$, fuzzy edge indicator function is much smaller at the region of edges or boundaries than at the homogeneous region. Then at the non-homogeneous region( $\theta(I) \rightarrow 0$) the proposed approach is less smooth while at the homogeneous region ($\theta(I) \rightarrow 1$) the proposed approach is more smooth. The proposed fuzzy edge indicator is capable to classify the noisy pixel from edges or boundaries based on the hesitation degree which is also called as the intuitionistic fuzzy index. The IFD function $F(I)$ have been computed as given in section \ref{sec:IFS}.
The above analysis leads us to propose a fuzzy edge detector driven convex total variation model for multiplicative speckle noise removal,
\begin{equation}\label{eq:new_energy_fuzzy}
\underset{I\in \Omega}{\text{argmin}}  \left\lbrace \int_{\Omega} \theta(I)|\nabla I|dx +\lambda \int_{\Omega}\left( I+I_0 \log\frac{1}{I}\right) dx\right\rbrace. 
\end{equation}  
The associated Euler-Lagrange equation of \eqref{eq:new_energy_fuzzy} then given by:
\begin{equation}\label{eq:steady_diffusion}
\left.\begin{aligned}
\text{div}\left(\theta(I)\frac{\nabla I}{|\nabla I|}\right)-\lambda \left( 1-\frac{I_0}{I}\right) &=0, \hspace{1.0cm} \text{in} \hspace{0.2cm} \Omega,\\
\frac{\partial I}{\partial n} &=0,\hspace{1.0cm} \text{in} \hspace{0.2cm} \partial \Omega.
\end{aligned}\right\}
\end{equation}
This system provides the steady-state solution and thus fails to produce the best effect. Hence, it is more meaningful to use the evolutionary version of \eqref{eq:steady_diffusion}. Therefore we are interested in the following telegraph total variation model
\begin{equation}\label{eq:final_fuzzy_telegrah}
\left.\begin{aligned}
I_{tt}+\gamma I_t &= \text{div}\left(\theta(I)\frac{\nabla I}{|\nabla I|}\right)-\lambda \left( 1-\frac{I_0}{I}\right),\hspace{0.3cm} \text{in} \hspace{0.2cm} \Omega_{T},  \\
\frac{\partial I}{\partial n} &=0,               \hspace{2.8cm} \text{in}
\hspace{0.2cm} \partial \Omega_{T},\\
I(x,0)&=I_0(x), \hspace{0.1cm}   I_t(x,0)=0,	\hspace{0.2cm} \text{in} \hspace{0.2cm}\Omega.
\end{aligned}\right\}
\end{equation}

\subsection{The advantages and novelty of the proposed model:}
The proposed approach possesses the following analogous advantages and novelty:
\begin{enumerate}
\item[$\bullet$] The energy functional \eqref{eq:new_energy_fuzzy} is globally convex and therefore the associated variational problem has a unique minimizer \cite{dong2013convex}.
\item[$\bullet$] Since the proposed model is a parabolic-hyperbolic PDE, it enables us to do edge preservation and image enhancement \cite{ratner2007image}. In  \cite{ratner2007image}, parabolic-hyperbolic PDE has been used only to remove the additive Gaussian noise. So the proposed model can remove the multiplicative speckle noise and preserve the significant features and structures of filtered images.
\item[$\bullet$] The fuzzy nature of speckle noise and edges is studied through the IFD function. In this fuzzy edge detection process, each noisy pixel in an image is mapped into different classes such as homogeneous, details and edges, using fuzzy set theory.
\item[$\bullet$] Based on all these inferences, the proposed model \eqref{eq:final_fuzzy_telegrah} with a fuzzy edge detector based filtering approach enables us for detail edge and structure preservation. 
\item[$\bullet$] To the best of our knowledge, the present work marks the first step towards the use of telegraph total variation based model with fuzzy edge indicator function for multiplicative speckle noise removal.
\end{enumerate}
\section{Existence and uniqueness of weak solution}
\label{sec:analysis}
In this section, we study the existence and uniqueness of weak solution of the regularized version of the proposed model \eqref{eq:final_fuzzy_telegrah}.
Consider the regularized model as
\begin{align}
\dfrac{\partial^2 I}{\partial t^2} +\gamma\dfrac{\partial I}{\partial t}= \text{div} \left( \dfrac{\theta(I_\xi)}{1+|\nabla G_{\xi}\ast I|}
\nabla I\right)
-\lambda \left(1-\dfrac{I_0}{I} \right)\,,&  \hspace{0.3cm} \text{in}\,\,\, \Omega_T\,, \label{maina} \\
\label{mainb}
\dfrac{\partial I}{\partial n}=0\,,& \hspace{0.3cm} \text{on}\,\,\, \partial \Omega_{T}\,,\\
I(x,0)=I_0(x)\,, \hspace{0.2cm} I_t(x,0)=0\,,& \hspace{0.3cm} \text{in}\,\,\, \Omega\,,\label{mainc}
\end{align}
where $I_\xi=G_\xi\ast I$.
Since the problem \eqref{maina}-\eqref{mainc} is nonlinear, we first consider the linearized problem, and then use Schauder’s
fixed-point theorem  \cite{LCEvans1998} to show the existence of a weak solution. For simplicity we choose all the constants involved in the equations \eqref{maina}-\eqref{mainc} equals to 1.
\subsection{Technical framework $\&$ statement of the main result:}
Throughout this paper, we use the letters $C$, $K$ etc to denote various generic constants. There are
situations where constants may change from line to line, but the notation is kept unchanged so long as it
does not impact the central idea.

We denote by $H^k(\Omega)$, $k$ is a positive integer, the set of all functions $I: \Omega \rightarrow \R $ such that $I$ and its distributional derivatives 
$\frac{\partial^m I}{\partial x^m}$
of order $|m|=\sum_{j=1}^{2}m_j \leq k$ all belongs to $L^2(\Omega)$. $H^k(\Omega)$ is a Hilbert space endowed with the norm
\begin{align*}
 ||I||_{H^k(\Omega)}=\Bigg(\underset{|m|\leq k }{\sum}\int_{\Omega} \Big|\dfrac{\partial^m I}{\partial x^m}\Big|^2dx \Bigg)^{1/2}\,.
 \end{align*}
For any Banach space $(X,\|\cdot\|_{X})$, we denote by $L^{p}(0,T; X)$, $p>1$, the set of all measurable functions $I:[0,T]\goto X$ such that 
$$\|I\|_{L^p(0,T; X)}:=\Bigg(\int_{0}^{T} || I(t)||^p_{X}\,dt \Bigg)^{1/p} < \infty.$$
Similarly, $L^{\infty}(0,T;X)$ denotes the set of all functions $I$ such that for a.e.~$t\in (0,T)$, $I(t)\in X$, and 
\begin{align*}
||I||_{L^{\infty}(0,T;X)}&=\underset{0<t<T }{\text{ess sup}}||I(t)||_{X}
=\text{inf}\Big\{ M: ||I(t)||_{X}\leq M,\hspace{0.1cm} \text{ a.e.~on (0,T)} \Big\}
 < \infty\,.
\end{align*}
We denote by $H^{1}(\Omega)'$ the dual of $H^{1}(\Omega)$. For any $f\in H^{1}(\Omega)'$, we define a norm as
$$||f||_{H^1(\Omega)'}= \Big\{ \sup\,\langle f,u \rangle :u \in H^1(\Omega)\,,\,\,\,||u||_{H^1(\Omega)} \leq 1   \Big\} .$$
Let us introduce the solution space $W$ of the problem \eqref{maina}-\eqref{mainc}:
\begin{align*}
\begin{split}
W=\Bigg\{ w\in L^{\infty}(0,T;H^{1}(\Omega)),\dfrac{\partial w}{\partial t} \in L^{\infty}(0,T;L^2(\Omega)),
 \dfrac{\partial^2 w}{\partial t^2} \in L^{2}(0,T;(H^{1}(\Omega))'): \\
0<\alpha=\underset{x \in \Omega}{\text{inf}}I_0(x) \leq w(x,t) \leq \underset{x \in \Omega}{\text{sup}}I_0(x)=\beta\,
 \quad \text{for a.e.}\, (x,t)\in \Omega\times (0,T)  \Bigg\}.
\end{split}
\end{align*}
Obviously, $W$ is a Banach space equipped with the norm
\begin{align*}
\begin{split}
\left\Vert w \right\Vert_{W}=\left\Vert w \right\Vert_{L^{\infty}(0,T;H^{1}(\Omega))} + \left\Vert \dfrac{\partial w}{\partial t}\right\Vert_{L^{\infty}(0,T;L^2(\Omega))} 
 + \left\Vert\dfrac{\partial^2 w}{\partial t^2}\right\Vert_{L^{2}(0,T;H^{1}(\Omega)')}.
\end{split}
\end{align*}
\begin{defi}
 A function $I$ is called a weak solution of the problem \eqref{maina}-\eqref{mainc}, if $I \in W$ and satisfies \eqref{maina}
 in the sense of distributions, i.e., for a.e. $t\in (0,T)$, there holds
\begin{align*}
\Big \langle \dfrac{\partial^2 I }{\partial t^2}, \phi   \Big \rangle + {\displaystyle \int_{\Omega}}
\Big(  \dfrac{\partial I}{\partial t} \phi +  \dfrac{\theta(I_\xi)}{1+|\nabla G_{\xi}\ast I|}\nabla I\cdot \nabla \phi  \Big)\,dx 
=-\displaystyle \int_{\Omega} \Big(1-\dfrac{I_0}{I} \Big)\phi\, dx\,, \quad \forall\,\phi
\in H^1(\Omega)\,, 
\end{align*}
along with the conditions \eqref{mainb} and \eqref{mainc}. 
\end{defi}
As we mentioned, our aim is to establish the existence and uniqueness of weak solutions of the underlying problem \eqref{maina}-\eqref{mainc}, and we will do so under the following assumptions:
\begin{Assumptions}
 \item \label{A1}  The initial data $I_0$ is an $H^2$-valued function satisfying
 \begin{align*}
  0< \alpha:=\inf_{x\in \Omega} I_0(x); \quad \beta:= \sup_{x\in \Omega} I_0(x) < \infty\,.
 \end{align*}

\item \label{A2} The function $\theta:\R \goto \R$ is a positive, bounded, Lipschitz function. More precisely, there exist $\delta, C_{\theta}>0$ such that 
\begin{align*}
 \begin{cases}
  \delta \le \theta(\cdot) \le 1\,, \\
  \big| \theta(x)-\theta(y)\big| \le C_{\theta}|x-y|\,, \quad \forall\, x,y \in \R\,.
 \end{cases}
\end{align*}
\end{Assumptions}

We are now ready to state the main results of this paper.

\begin{thm}\label{thm:existence-uniqueness}
Let the assumptions \ref{A1}-\ref{A2} be true. Then the problem \eqref{maina}-\eqref{mainc} admits one and only one weak solution. 
\end{thm}
\subsection{Linearized problem  $\&$ existence of weak solution:} For any fixed $ w \in W $, first we consider the following linearized problem :
\begin{align}
&\Big \langle \dfrac{\partial^2 I_w }{\partial t^2}, \phi  \Big \rangle + \large \int_{\Omega} \Big(  \dfrac{\partial I_w}{\partial t} \phi +  g_w \nabla I_w \cdot\nabla \phi  \Big)\,dx 
= -\int_{\Omega} \Big(1-\dfrac{I_0}{w} \Big)\phi \,dx, \hspace{0.2cm} \forall \phi \in H^1, \label{linmaina} \\
&I_w\left(x,0\right)=I_0(x)\,,\ \dfrac{\partial I_w}{\partial t}\left(x,0\right)=0  \,,\,\, \text{in} \,\,\,\Omega\,, \label{linmainb}
\end{align}
with
\begin{align}\label{boundw}
\left\Vert w \right\Vert_{L^{\infty}(0,T;H^{1}(\Omega))} + \left\Vert \dfrac{\partial w}{\partial t}\right\Vert_{L^{\infty}(0,T;L^2(\Omega))} \leq C\left\Vert I_0 \right\Vert_{H^1},
\end{align}
where $C$ is a positive constant, and the function $g_w(x,t)$ is defined as
\begin{align*}
 g_w(x,t)=\dfrac{\theta(w_\xi(x,t))}{1+|\nabla G_{\xi}\ast w(x,t)|}\,.
\end{align*}
In view of the property of convolution, the assumption \ref{A2}, and \eqref{boundw}, the following inequalities hold:
\begin{align}
\kappa: =\dfrac{\delta}{1+C\left(G_{\xi},\left\Vert I_0 \right\Vert_{H^1}\right)} \leq g_w \leq 1\,; \quad 
 \left\vert \dfrac{\partial g_w}{\partial t}  \right\vert  \le  C_1\,, \label{bound:g_w}
\end{align}
for some constant $ C_1>0 $, depending only on $\theta, \Omega, G_\xi \,\, \text{and} \,\,\left\Vert I_0 \right\Vert_{H^1}$. Indeed, since 
\begin{align*}
1+ |\nabla G_{\xi}\ast w| & \le 1 + \|\nabla G_{\xi}\|_{L^2}\|w\|_{L^\infty(0,T; L^2(\Omega))}\\
&\le 1+  C_{\xi} C\left\Vert I_0 \right\Vert_{H^1} \\
&\equiv 1+ C\left(G_{\xi},\left\Vert I_0 \right\Vert_{H^1}\right),
\end{align*}
we see that 
\begin{align*}
 \dfrac{1}{1+C\left(G_{\xi},\left\Vert I_0 \right\Vert_{H^1}\right)} \leq \dfrac{1}{1+|\nabla G_{\xi}\ast w|} \leq 1\,.
\end{align*}
Therefore, thanks to the assumption \ref{A2}, we obtain
\begin{align*}
 \kappa:= \dfrac{\delta}{1+C\left(G_{\xi},\left\Vert I_0 \right\Vert_{H^1}\right)}\le \dfrac{\theta(w_\xi)}{1+|\nabla G_{\xi}\ast w|}:=g_w(x,t)\le 1\,.
\end{align*}
To see the second inequality of \eqref{bound:g_w}, notice that
\begin{align*}
 \left\vert \dfrac{\partial g_w}{\partial t} \right\vert & \leq  C_{\theta} |G_\xi \ast w_t| + | \nabla G_{\xi}\ast w_t| \\
 & \le C_{\theta}\|G_\xi\|_{L^2(\Omega)} \|w_t\|_{L^\infty(0,T; L^2(\Omega))} \\
 & \hspace{1cm}+ \|\nabla G_\xi\|_{L^2(\Omega)}\|w_t\|_{L^\infty(0,T; L^2(\Omega))} \\
 & \le C_1(\theta, \Omega, G_{\xi}, \|I_0\|_{H^1(\Omega)})\,.
\end{align*}
Since $g_w(x,t)$ satisfies \eqref{bound:g_w}, one can apply classical Galerkin method \cite{LCEvans1998} to show that 
the linearized problem \eqref{linmaina}-\eqref{linmainb} has a unique weak solution $I_w \in W$.

\subsubsection{Energy Estimates} Note that $\dfrac{\partial I_w}{\partial t} \in L^\infty(0,T; H^1(\Omega))$. Taking $\phi=\dfrac{\partial I_w}{\partial t}$ in \eqref{linmaina}, integrating by parts, we have 
\begin{align*}
&\frac{1}{2}\dfrac{d}{dt}\|\dfrac{\partial I_w}{\partial t}\|_{L^2}^2 +\|\dfrac{\partial I_w}{\partial t}\|_{L^2}^2
+ \large\int_{\Omega} g_w\nabla I_w\cdot \nabla \big(\dfrac{\partial I_w}{\partial t}\big)\, dx \nonumber \\
& ={\large\int_{\Omega}}\Big(\dfrac{I_0}{w}-1 \Big) \dfrac{\partial I_w}{\partial t}\,dx \le 
 \dfrac{1}{2}\big\|\dfrac{I_0}{w}-1 \big\|_{L^2}^2 + \dfrac{1}{2}\|\dfrac{\partial I_w}{\partial t}\|_{L^2}^2\,. 
\end{align*}
Note that, thanks to integration by parts formual and \eqref{bound:g_w},
\begin{align*}
\int_{\Omega} g_w\nabla I_w\cdot\nabla \frac{\partial I_w}{\partial t}\, dx 
&= \frac{1}{2}\dfrac{d}{dt}\int_{\Omega} g_w |\nabla I_{w}|^2\, dx 
-\frac{1}{2}\int_{\Omega}\frac{\partial g_w}{\partial t}| \nabla I_{w}|^2\, dx\\
&\geq  \frac{1}{2}\dfrac{d}{dt}\int_{\Omega} g_w|\nabla I_{w}|^2 \,dx
-\frac{C_1}{2}\|\nabla I_w(t)\|_{L^2(\Omega)}^2\,.
\end{align*}
Again, thanks to \eqref{boundw} and \ref{A1}, we get that
\begin{align*}
\big\|\dfrac{I_0}{w}-1 \big\|_{L^2}^2  &\leq \dfrac{1}{\alpha^2}\left\Vert I_0-w \right\Vert^{2}_{L^2} \\
&\leq \dfrac{2}{\alpha^2}\left( \left\Vert I_0 \right\Vert^{2}_{H^1} + \left\Vert w \right\Vert^{2}_{H^1} \right) \\
&\le \dfrac{2}{\alpha^2}\left(1+C^2 \right)\left\Vert I_0 \right\Vert^2_{H^1}.
 \end{align*}
Combining the above two estimates, we get  
\begin{align}
&\frac{d}{dt}\Big[ \| \frac{\partial I_{w}}{\partial t}\|_{L^2(\Omega)}^2 + \int_{\Omega} g_w |\nabla I_{w}|^2\, dx\Big] \nonumber \\
&   \le \dfrac{1+C^2}{\alpha^2}\left\Vert I_0 \right\Vert^2_{H^1} + C_1\|\nabla I_w(t)\|_{L^2}^2 + \| \frac{\partial I_{w}}{\partial t}\|_{L^2(\Omega)}^2  \notag \\
&   \equiv C_2 + C_1\|\nabla I_w(t)\|_{L^2}^2 + \| \frac{\partial I_{w}}{\partial t}\|_{L^2(\Omega)}^2 \,. \label{esti:1}
\end{align}
Thanks to the lower bound of $g_w$ as in \eqref{bound:g_w}, we observe that
\begin{align}
 \|\nabla I_w(t)\|_{L^2(\Omega)}^2 \le \frac{1}{\kappa}  \int_{\Omega} g_w |\nabla I_{w}|^2\, dx\, \label{esti:gradient-inters-gw}
\end{align}
and hence, we obtain from \eqref{esti:1}
\begin{align*}
&\frac{d}{dt}\Big[ \| \frac{\partial I_{w}}{\partial t}\|_{L^2(\Omega)}^2 + \int_{\Omega} g_w |\nabla I_{w}|^2\, dx\Big] 
 \le C_2 + C_3 \Big(\| \frac{\partial I_{w}}{\partial t}\|_{L^2(\Omega)}^2 + \int_{\Omega} g_w |\nabla I_{w}|^2\, dx\Big)
\end{align*}
where $C_3=\text{max}\left\{1,\dfrac{C_1}{\kappa} \right\}.$ An application of Gronwall's lemma gives: for a.e. $t\in (0,T]$
\begin{align*}
&\big\|\dfrac{\partial I_w(t)}{\partial t}\big\|_{L^2}^2 + \int_{\Omega}g_w(x,t)|\nabla I_w(x,t)|^2\,dx \le e^{C_3t} \left(C_4 + tC_2 \right),
\end{align*}
where $ \,\,C_4=\big\|\dfrac{\partial I_w(0)}{\partial t}\big\|_{L^2}^2 + \displaystyle \int_{\Omega} g_w(x,0)|\nabla I_w(x,0)|^2\,dx$.
Moreover, in view of \eqref{esti:gradient-inters-gw}, 
\begin{align*}
\left\Vert \nabla I_w(t) \right\Vert^2_{L^2} \le \dfrac{1}{\kappa} e^{C_3t} \left(C_4 + tC_2 \right).
\end{align*}
Thus, for a.e. $t\in (0,T]$, 
\begin{align}\label{bound_I_t_nabla_I}
\big\|\dfrac{\partial I_w(t)}{\partial t}\big\|_{L^2}^2  + \|\nabla I_w(t)\|_{L^2}^2 \leq M_1 e^{C_3t} \left(C_4 + tC_2 \right),
\end{align}
where $M_1=\text{max}\left\{ \dfrac{1}{\kappa},1 \right\}$.\newline Since $I(x,t)=I(x,0)+ \displaystyle \int_{0}^{t}\dfrac{\partial I}{\partial s}ds$, we have, thanks to Young's inequality and \eqref{bound_I_t_nabla_I},
also we have
\begin{align} \label{boundIL2}
\left\Vert I_w(t) \right\Vert^2_{L^2} &\leq 2\left\Vert I_0 \right\Vert^2_{H^1} + 2T\int_{0}^{t}\big\|\dfrac{\partial I_w(s)}{\partial s}\big\|_{L^2}^2\, ds \nonumber \\
&\le 2\left\Vert I_0 \right\Vert^2_{H^1} + 2T^2  M_1 e^{C_3T} \left(C_4 + TC_2 \right)\,.
\end{align}
We combine \eqref{bound_I_t_nabla_I} and \eqref{boundIL2} to conclude 
\begin{align}\label{boundIH1boundItL2}
\left\Vert I_w \right\Vert_{L^{\infty}(0,T;H^{1}(\Omega))} + \big\|\dfrac{\partial I_w}{\partial t}\big\|_{L^{\infty}(0,T;L^2(\Omega))} \leq C_5\left\Vert I_0 \right\Vert_{H^1}.
\end{align}
Now choose $\phi \in H^{1}(\Omega)$ with $||\phi||_{H^{1}(\Omega)}\leq 1$ in \eqref{linmaina}, and use Cauchy-Schwarz inequality along with \eqref{boundIH1boundItL2} to obtain 
\begin{align*}
\begin{split}
\Big|\Big \langle \dfrac{\partial^2 I_{w} }{\partial t^2}, \phi \Big \rangle \Big|
&\leq  \Big(\big\|\dfrac{\partial I_{w}}{\partial t}\big\|_{L^2} +\left\vert g_w \right\vert\left\Vert \nabla I_{w}(t)\right\Vert_{L^2} + \big\|1-\dfrac{I_0}{w}\big\|_{L^2}\Big)\|\phi\|_{H^1}  \\
&\leq  \Big\{ 2 C_5 +\sqrt{\dfrac{2}{\alpha^2}(1+C^2)}\Big\} \|I_0\|_{H^1(\Omega)} \left\Vert\phi\right\Vert_{H^1}. 
\end{split}
\end{align*}

Hence, by the definition of norm in $H^{1}(\Omega)'$, we infer that $
\Big|\Big|\frac{\partial^2 I_{w} }{\partial t^2}\Big|\Big|_{H^1(\Omega)^\prime} \leq C_6\|I_0\|_{H^2(\Omega)} $. Squaring and integrating over $(0,T)$, we obtain

\begin{equation}\label{bounddel2IdelT2}
\int_{0}^{T}\Big|\Big|\dfrac{\partial^2 I_{w}(t) }{\partial t^2}\Big|\Big|^{2}_{H^{1}(\Omega)'}\, dt \leq C_6\int_{0}^{T}||I_0||^2_{H^1(\Omega)}\,dt\,.
\end{equation}

\subsubsection{Passing to the limit}
From \eqref{boundIH1boundItL2} and \eqref{bounddel2IdelT2}, we introduce the subspace $W_0$ of $W$ defined by
\begin{align*}
W_0=\Big\{ w \in W:\,||w||_W \leq C\|I_0\|_{H^1}^2, \hspace{0.05cm}
 w(0)=I_0\,,  \dfrac{\partial w(0)}{\partial t}=0   \Big\}\,.
\end{align*}
Moreover, one can prove that $W_0$ is a non-empty, convex and
weakly compact subset of $W$. Consider a mapping
\begin{align*}
 \mathcal{P}:~ & W_0 \goto W_0 \\
 & w\mapsto I_w\,.
\end{align*}
In order to use Schauder's fixed-point theorem on $\mathcal{P}$, we need to prove only that the mapping $\mathcal{P}:w \rightarrow I_w $ is weakly continuous from $W_0$ into $W_0$. Let
 $w_k$ be a sequence that converges weakly to some $w$ in $W_0$ and let $I_k = I_{w_k}$. We have to show that $\mathcal{P}(w_k):= I_k$ converges weakly
 to $\mathcal{P}(w): = I_w$.
 \vspace{.1cm}
 
From the classical results of compact inclusion in Sobolev spaces \cite{raadams1975}, we can extract subsequences of $\{w_k\}$ and $\{I_k\}$ still denoted by $\{w_k\}$ and $\{I_k\}$ respectively such that for some $I\in W_0$, we have, as $k\goto \infty$

\begin{align*}
\begin{cases}
 w_{k} \longrightarrow  w \hspace{0.2cm} \text{in} \hspace{0.2cm} L^2(0,T;L^2(\Omega)) \hspace{0.2cm} \text{ and a.e. on } \hspace{0.2cm}  \Omega \times (0,T),\\
 \displaystyle  \dfrac{1}{w_k}  \longrightarrow \dfrac{1}{w} \hspace{0.2cm} \text{in} \hspace{0.2cm} L^{2}(0,T;L^2(\Omega)) \hspace{0.2cm}  \text{and a.e. on} \hspace{0.2cm}   \Omega \times (0,T),\\
 G_{\xi}\ast w_k  \longrightarrow G_{\xi}\ast w  \hspace{0.2cm} \text{in} \hspace{0.2cm}  L^2(0,T;L^2(\Omega))\\
  \hspace{0.2cm}  \text{and a.e. on} \hspace{0.2cm}   \Omega \times (0,T),\\
\theta(G_{\xi}\ast w_k) \longrightarrow \theta(G_{\xi}\ast w) \hspace{0.2cm} \text{in} \hspace{0.2cm}  L^2(0,T;L^2(\Omega)) \\
  \text{and a.e. on} \hspace{0.2cm}   \Omega \times (0,T),\\
\dfrac{\partial G_{\xi}}{\partial x_n}\ast w_k  \longrightarrow \dfrac{\partial G_{\xi}}{\partial x_n}\ast w  \hspace{0.2cm} \text{in} \hspace{0.2cm}  L^2(0,T;L^2(\Omega))\\
 \hspace{0.2cm}  \text{and a.e. on} \hspace{0.2cm}   \Omega \times (0,T), n=1,2,\\
\dfrac{1}{1 + |\nabla G_{\xi}\ast w_k|}  \longrightarrow \dfrac{1}{1 + |\nabla G_{\xi}\ast w|}  \hspace{0.2cm} \text{in} \hspace{0.2cm}  L^2(0,T;L^2(\Omega))\\ \hspace{0.2cm}  \text{and a.e. on} \hspace{0.2cm}   \Omega \times (0,T),\\
\dfrac{\theta(G_{\xi}\ast w_k)}{1 + |\nabla G_{\xi}\ast w_k|}  \longrightarrow \dfrac{\theta(G_{\xi}\ast w)}{1 + |\nabla G_{\xi}\ast w|}  \hspace{0.2cm} \text{in} \hspace{0.2cm}  L^2(0,T;L^2(\Omega))\\
 \hspace{0.2cm}  \text{and a.e. on} \hspace{0.2cm}   \Omega \times (0,T),\\
 \displaystyle I_{k} \longrightarrow  I \hspace{0.2cm} \text{weakly} * \text{in} \hspace{0.2cm} L^{\infty}(0,T;H^1(\Omega)),\\
 \dfrac{\partial I_k}{\partial t} \longrightarrow  \dfrac{\partial I}{\partial t}  \hspace{0.2cm} \text{weakly} * \text{in} \hspace{0.2cm} L^{\infty}(0,T;L^2(\Omega)),\\
 \dfrac{\partial^2 I_k}{\partial t^2} \longrightarrow  \dfrac{\partial^2 I}{\partial t^2}  \hspace{0.2cm} \text{weakly} * \text{in} \hspace{0.2cm} L^{2}(0,T;H^1(\Omega)^{\prime}),\\
 I_{k} \longrightarrow  I \hspace{0.2cm} \text{in} \hspace{0.2cm} L^{2}(0,T;L^2(\Omega)),\\
 \dfrac{\partial I_k}{\partial x_k} \longrightarrow  \dfrac{\partial I}{\partial x_k}  \hspace{0.2cm} \text{weakly} * \text{in} \hspace{0.2cm} L^{\infty}(0,T;L^2(\Omega)).\\
\end{cases}
\end{align*}

The above convergence allow us to pass to the limit in the problem \eqref{linmaina} and obtain $I=\mathcal{P}(w)$.  Moreover, since the solution of \eqref{linmaina} is unique, the whole
sequence $I_k=\mathcal{P}(w_k)$ converges weakly in $W_0$ to $I=\mathcal{P}(w)$. Hence $\mathcal{P}$ is weakly continuous. Consequently, thanks to the Schauder fixed
point theorem, there exists $w \in W_0$ such that $w=\mathcal{P}(w)=I_w$.  Thus, the function $I_w$ solves the problem \eqref{maina}-\eqref{mainc}.
\subsection{Uniqueness of weak solution:}
Following the idea as in \cite{LCEvans1998}, we prove the uniqueness of weak solutions of the underlying problem \eqref{maina}-\eqref{mainc}. Let $I_{1}$ and $I_{2}$ be two weak solutions of \eqref{maina}-\eqref{mainc}.
Then for almost every $t\in(0,T)$, we have
\begin{equation}\label{eq:i-uniqueness}
\begin{aligned}
\dfrac{\partial^2 I_{i}}{\partial t^2}+\dfrac{\partial I_{i}}{\partial t}-\text{div}\left( g_{i} \nabla I_{i}\right) & =-\Big(1-\dfrac{I_0}{I_i} \Big),\, \hspace{0.5cm} \text{in} \,\,\,  \Omega_T\,,\\
\dfrac{\partial I_{i}}{\partial n} & =0,\, \hspace{2.0cm} \text{on} \,\,\,  \partial \Omega_T\,,\\
I_{i}(x,0)=I_0(x)\,,\,\,\, I_t(x,0)&=0,\, \hspace{2.0cm} \text{in} \,\,\,  \Omega,\, 
\end{aligned}
\end{equation}
where $g_i= \dfrac{\theta(G_\xi \ast I_i)}{1+|\nabla G_{\xi}\ast I_i|},i=1,2.$ Let $\tilde{I}:= I_1 - I_2$. Then, by subtracting for $i=1,2$, we obtain, from \eqref{eq:i-uniqueness}
\begin{align}
\label{unique4}
&\dfrac{\partial^2 \tilde{I}}{\partial t^2}+\dfrac{\partial \tilde{I}}{\partial t}-\text{div}\big(g_{1} \nabla \tilde{I}\big)
=\text{div}\left( \left(g_{1}-g_{2}\right) \nabla I_{2}\right)- \dfrac{I_0}{I_1I_2}\tilde{I}\,,
 \hspace{0.5cm}\text{in} \,\,\,\Omega_T\,,\\
\label{unique5}
&\dfrac{\partial \tilde{I} }{\partial n}=0\,, \hspace{3.8cm} \text{on}\,\, \partial \Omega_T\,,\\
\label{unique6}
&\tilde{I}(x,0)=0\,,\,\,\,\dfrac{\partial \tilde{I}}{\partial t}(x,0)=0\,, \hspace{1.1cm} \text{in} \,\,\, \Omega.
\end{align}
It suffices to show that $ \tilde{I} \equiv 0$ . To verify this, fix $ 0 < s < T$, and set for $i=1,2$, 
\begin{align}\label{relationvi}
v_{i}(.,t)= \begin{cases}
\displaystyle \int_{t}^{s} I_{i}(.,\tau)d\tau, \hspace{0.5cm} 0<t\leq s\,, \\ 
 0 \hspace{2.5cm} s \leq t < T\,.
\end{cases}
\end{align}
Note that, for $t\in (0,T)$,
\begin{align}\label{eq:fact-1}
 \begin{cases}
 \dfrac{\partial v_i}{\partial t}(x,t)=-I_i(x,t) \quad i=1,2\,, \\
  v_{i}(\cdot,t) \in H^1(\Omega)\,, \,\,\, \dfrac{\partial v_{i}  }{\partial n}=0\, \text{on}\,\, \partial \Omega\,\,\\
\text{in the sence of distribution}.
 \end{cases}
\end{align}
Set $v=v_1-v_2$. Then $v(\cdot, s)=0$. 
Multiplying \eqref{unique4} by $v$, integrating over $\Omega \times (0,s)$ and using \eqref{unique5} and \eqref{unique6} along with the integration by parts formula, we obtain
\begin{align*}
&\int_{0}^{s}\int_{\Omega} \Big( - \dfrac{\partial\tilde{I}}{\partial t}\dfrac{\partial v}{\partial t} - \tilde{I}\dfrac{\partial v}{\partial t}+g_{1} \nabla \tilde{I} \cdot \nabla v \Big)\,dx\,dt 
= -\int_{0}^{s}\int_{\Omega}  \big(g_{1}-g_{2}\big) \nabla I_{2}\cdot \nabla v\,dx\,dt-\int_{0}^{s}\int_{\Omega} \dfrac{I_0}{I_1I_2} \tilde{I} v\,dx\,dt.
\end{align*}
 We use \eqref{eq:fact-1} in the above equality, and then use Cauchy-Schwarz inequality along with the fact that $\frac{I_0}{I_1 I_2}\le \frac{\beta}{\alpha^2}$ to get
\begin{align*}
&\frac{1}{2}\int_{0}^{s}\int_{\Omega} \dfrac{\partial}{\partial t}  |\tilde{I}|^2 dxdt+\int_{0}^{s}\int_{\Omega}|\tilde{I}|^2\,dx\,dt
-\int_{0}^{s}\int_{\Omega}g_{1}
\dfrac{\partial \nabla v}{\partial t}\cdot \nabla v\,dx\,dt\\
&= -\int_{0}^{s}\int_{\Omega}  \big(g_{1}-g_{2}\big) \nabla I_{2}\cdot \nabla v\,dx\,dt-\int_{0}^{s}\int_{\Omega} \dfrac{I_0}{I_1I_2} \tilde{I} v\,dx\,dt \\
&\leq -\int_{0}^{s}\int_{\Omega}(g_{1}-g_{2}) \nabla I_{2}\cdot \nabla v\,dx\,dt
+\dfrac{\beta}{2\alpha^2}\int_{0}^{s}\int_{\Omega} |\tilde{I}|^2\,dx\,dt
+\dfrac{\beta}{2\alpha^2}\int_{0}^{s}\int_{\Omega} |v|^2\,dx\,dt \\
& \le \int_0^s \|(g_1-g_2)(t)\|_{L^\infty(\Omega)} \|\nabla I_2(t)\|_{L^2}\|\nabla v(t)\|_{L^2}\,dt
 + \dfrac{\beta}{2\alpha^2}\int_{0}^{s}\|\tilde{I}(t)\|_{L^2}^2\,dt
+\dfrac{\beta}{2\alpha^2}\int_{0}^{s}\|v(t)\|_{L^2}^2\,dt\,.
\end{align*}
Now using the fact that
\begin{align*}
 g_{1}\dfrac{\partial \nabla v}{\partial t}\cdot \nabla v &= \frac{1}{2} \frac{\partial}{\partial t}(g_1 |\nabla v|^2)-\frac{1}{2} \frac{\partial g_1}{\partial t}|\nabla v|^2\,, \\
 \nabla v(x, s)&=0\,,
\end{align*}
and \eqref{unique6}, we have
\begin{align}\label{unique9}
&\frac{1}{2}\|\tilde{I}(s)\|_{L^2}^2+\int_{0}^{s}\|\tilde{I}(t)\|_{L^2}^2\,dt  + \frac{1}{2}\int_{\Omega} g_{1}(x,0) |\nabla v(x,0)|^2\, dx \nonumber  \\
&\leq \Big|-\frac{1}{2}\int_{0}^{s}\int_{\Omega} |\nabla v|^2 \dfrac{\partial g_1}{\partial t}\, dx\,dt\Big| 
 + \int_0^s \|(g_1-g_2)(t)\|_{L^\infty(\Omega)} \|\nabla I_2(t)\|_{L^2}\|\nabla v(t)\|_{L^2}\,dt \notag \\
 & \hspace{2cm} +\dfrac{\beta}{2\alpha^2}\int_{0}^{s}\|\tilde{I}(t)\|_{L^2}^2\,dt
+\dfrac{\beta}{2\alpha^2}\int_{0}^{s}\|v(t)\|_{L^2}^2\,dt\,.
\end{align}
As seen in the proof of existence, there exist positive constants $C_7$ and $C_8$ such that
\begin{align*}
C_7 \leq g_{i}(x,t)\leq 1\,,\hspace{1cm}  \left\vert \dfrac{\partial g_i(x,t)}{\partial t} \right\vert\leq  C_8\,,\\
\hspace{0.1cm}  \text{for a.e.} (x,t)\in \Omega_T \,\text{and}\, i=1,2.
\end{align*}
Moreover, one can use property of convolution along the stated assumptions on $\theta$ to show that 
$$||(g_{1}-g_{2})(t)||_{L^{\infty}(\Omega)} \leq C_{9}||\tilde{I}(t)||_{L^{2}(\Omega)} $$ holds for some constant $C_{9}>0$. Thus, using the above estimates in  \eqref{unique9}, we have
\begin{align}\label{unique14}
\frac{1}{2}\|\tilde{I}(s)\|_{L^2}^2+\int_{0}^{s}\|\tilde{I}(t)\|_{L^2}^2\,dt  + \frac{C_7}{2} \|\nabla v(0)\|_{L^2}^2
&\le \frac{C_8}{2}\int_{0}^{s} \|\nabla v(t)\|_{L^2}^2\,dt 
 + C_9 \|I_2\|_{L^\infty(0,T; H^1)}\int_{0}^{s} \|\tilde{I}(t)\|_{L^2}\|\nabla v(t)\|dt \nonumber \\
& \hspace{1cm} +\dfrac{\beta}{2\alpha^2}\int_{0}^{s}\|\tilde{I}(t)\|_{L^2}^2\,dt
 +\dfrac{\beta}{2\alpha^2}\int_{0}^{s}\|v(t)\|_{L^2}^2\,dt \notag \\
& \le  C \Big( \int_{0}^{s}\|\tilde{I}(t)\|_{L^2}^2\,dt + \int_{0}^{s}\|v(t)\|_{H^1}^2\,dt \Big)\,.
\end{align}
Observe from \eqref{relationvi} that
\begin{align}\label{extra1}
\|v(0)\|_{L^2}^2 = \big\|\int_0^s \tilde{I}(t)\,dt\big\|_{L^2}^2 \le T \int_0^s \|\tilde{I}(t)\|_{L^2}^2\,dt\,.
\end{align}
Using \eqref{extra1} in \eqref{unique14}, we have
\begin{align}\label{unique15_1}
\frac{1}{2}\|\tilde{I}(s)\|_{L^2}^2+\int_{0}^{s}\|\tilde{I}(t)\|_{L^2}^2\,dt  + \frac{C_7}{2} \|v(0)\|_{H^1}^2 
\leq C \Big( \int_{0}^{s}\|\tilde{I}(t)\|_{L^2}^2\,dt + \int_{0}^{s}\|v(t)\|_{H^1}^2\,dt \Big)\,.
\end{align}
Now let us write
\begin{align*}
w_{i}(.,t)&=\int_{0}^{t} I_{i}(.,\tau)d\tau\, ;\\
w(\cdot, t)&=(w_1-w_2)(\cdot,t)\,, \hspace{0.5cm} 0<t\leq T.
\end{align*}
In view of the above definition and \eqref{eq:fact-1}, we notice that
\begin{align*}
v(x,0)&= v(x,s) + \int_0^s \tilde{I}(x,t)\,dt 
= \int_0^s \tilde{I}(x,t)\,dt= w(x,s) \\
v(x,t)&= w(x,s)-w(x,t)\,, \quad 0<t\le s\,.
\end{align*}
Hence \eqref{unique15_1} reduces to
\begin{align}\label{unique16}
 \frac{1}{2}\|\tilde{I}(s)\|_{L^2}^2+\int_{0}^{s}\|\tilde{I}(t)\|_{L^2}^2\,dt  + \frac{C_7}{2} \|w(s)\|_{H^1}^2 
&\leq C \Big( \int_{0}^{s}\|\tilde{I}(t)\|_{L^2}^2\,dt + \int_{0}^{s}\|w(s)-w(t)\|_{H^1}^2\,dt \Big) \notag \\
& \le C  \int_{0}^{s}\|\tilde{I}(t)\|_{L^2}^2\,dt 
 + 2Cs \|w(s)\|_{H^1}^2 + 2C\int_0^s \|w(t)\|_{H^1}^2\,dt\,.
\end{align}
Now choose $T_1$ sufficiently small such that 
\begin{align*}
\frac{C_7}{2}-2T_1C>0.
\end{align*}
Then, for $0<s\leq T_1,$ we have, from \eqref{unique16}
\begin{align*}
\|\tilde{I}(s)\|_{L^2}^2 + \|w(s)\|_{H^1}^2 \le \tilde{C} \int_0^s\Big(\|\tilde{I}(t)\|_{L^2}^2 + \|w(t)\|_{H^1}^2\Big)\,dt\,, 
\end{align*}
for some constant $\tilde{C}>0$. 
Consequently, an application of Gronwall's lemma then implies $\tilde{I} \equiv 0$ on $[0,T_1]$.
Finally, we apply the same argument on the intervals $(T_1,2T_1]$, $(2T_1,3T_1],\ldots$ step by step, and eventually deduce that
$I_{1} = I_{2}$ on $(0,T)$.

\section{Numerical Implementation}
\label{sec:numerical}
The numerical discretization of the present model is required to process digital images. Hence, the numerical solution of \eqref{eq:final_fuzzy_telegrah} can be obtained using an iterative approach. We use an explicit finite difference scheme, to solve the model, which is taken as the most straightforward option in the literature. Also, we use a small time step to preserve the stability criterion \cite{jain2015iterative}.
The discrete explicit scheme is given as follows,\newline
(a). Let $\tau$ be the time step and $h$ the spatial step size. Denote $I^n_{i,j}=I(x_i,y_j,t_n)$ where $x_i=ih, \hspace{0.2cm} i=0,1,2...,N;$
$y_j=jh, \hspace{0.2cm} j=0,1,2...,M;$ $t_n=n\tau,\hspace{0.2cm}  n=0,1,2...$  where $M \times N$ is the size of the image.\\
(b). The symmetric boundary conditions are given as follows:
$I_{-1,j}^n=I_{0,j}^n,I_{N+1,j}^n=I_{N,j}^n, \hspace{0.2cm} I_{i,-1}^n=I_{i,0}^n,I_{i,M+1}^n=I_{i,M}^n.$	\\	
(c). The approximation of derivative terms using finite differences are given as follows:
\begin{center}
$
\begin{array}{lll}
\displaystyle\frac{\partial I}{\partial t} & \approx & \displaystyle\frac{I_{i,j}^{n+1}-I_{i,j}^n}{\tau},
\displaystyle\frac{\partial^2 I}{\partial t^2} \approx \displaystyle\frac{I_{i,j}^{n+1}-2I_{i,j}^n+I_{i,j}^{n-1}}{\tau ^2},\\[0.8cm]
\nabla_x I &\approx& \displaystyle\frac{I_{i+h,j}^n-I_{i-h,j}^n}{2h},
\nabla_y I \approx \displaystyle\frac{I_{i,j+h}^n-I_{i,j-h}^n}{2h}.
\end{array}
$
\end{center}
(d). The discretized version of the proposed filter \eqref{eq:final_fuzzy_telegrah} could be written as follows:
\begin{align}\label{discrete_system}
&(1+\gamma \tau)I_{i,j}^{n+1} \nonumber \\
&=(2+\gamma \tau)I_{i,j}^n -I_{i,j}^{n-1} 
+{\tau ^2} \Bigg[ \nabla_x \left( \theta(I_{i,j}^n) \frac{\nabla_x I_{i,j}^n}{\sqrt{(\nabla_x I_{i,j}^n)^2 + (\nabla_y I_{i,j}^n)^2}}\right) \nonumber \\
& + \nabla_y \left( \theta(I_{i,j}^n) \frac{\nabla_y I_{i,j}^n}{\sqrt{(\nabla_x I_{i,j}^n)^2 + (\nabla_y I_{i,j}^n)^2}}\right) \Bigg] 
-\tau ^2{\lambda}^n \left( 1-\frac{I_0}{I_{i,j}^n}\right),
\end{align}
where fuzzy edge indicator function can be calculated via,
$\theta(I_{i,j}^n)=1-F(I_{i,j}^n), \hspace{0.1cm}  F(I_{i,j}^n)=Div\_measure(i,j),  $
with the conditions, $I_{i,j}^0=I_0(ih,jh),  \hspace{0.2cm}  I_{i,j}^1=I_{i,j}^0.$
Through the above numerical discretization, we can obtain the solution at time T. Our numerical results depend on three parameters: the time step $\tau$, the damping coefficient $\gamma$ and the weight coefficient $\lambda$.
Apart from the numerical discretization of \eqref{eq:final_fuzzy_telegrah}, a convergence criterion is required to stop the elimination process. To reach our destination, we started with a corrupted image $I_0$ and used the system \eqref{discrete_system} repeatedly, resulting in a family of despeckled images ${I^p}$, which drafts the restored form of $I_0$. After sufficient iterations, changes between two consecutive iterations become redundant. To achieve the  convergence of the iterative processes, we used the  stopping criterion given below,
\begin{equation}\label{eq:4stopping}
\frac{{||I^{p+1}-I^p||^2_2}}{{||I^p||^2_2}}\leq \varepsilon,
\end{equation} 
where $\varepsilon > 0$ is a predefined threshold. In \eqref{eq:4stopping} $I^p$ and $I^{(p+1)}$ illustrate  the restored images  at the $p^{th}$ and ${(p+1)}^{th}$ iteration, respectively. For our simulations, we have used $\varepsilon \leq 10^{-4}$.

\section{Experiment Results and Discussion}
\label{sec:Results}
This section deals with qualitative and quantitative results which are described in two subsections. First, we validate the despeckling efficiency of the proposed model with the several existing PDE based models using standard synthetic and natural images. The numerical experiments using these digital images with ground truth information will enable us to quantify the efficiency of the despeckling algorithm. Whereas, the ability of the proposed approach has been investigated by using the real SAR images, which will illustrate the useful application of the proposed method for image processing.
\subsection{Setup and Parameters}
\label{sec:setup}
To see the despeckling ability of the proposed approach, we performed a large number of computational experiments using a group of natural images as well as real SAR images. For the study of despeckling, these natural images are initially corrupted by adding artificial multiplicative speckle noise with different noise level (look) ranging from $1$ to $33$ by using our MATLAB code. All the numerical results are computed under windows $7$ and MATLAB version $R2015b$ running on a desktop with an Intel Core $i5$ dual-core CPU at $2.53$ GHz with $4$ GB of memory. Despeckled images obtained by the proposed approach have been compared with the corresponding despeckling results of other discussed state-of-art methods.
In this process, the considered existing models are discretized using the same explicit numerical scheme as in the proposed model. The time step size ($\tau = 0.1$) and stopping criterion \eqref{eq:4stopping} are kept the same for each smoothing algorithm. Also, for fair and effective comparison, different parameters of considered models are optimized manually to obtain their best performance level.
\subsection{Image quality measurement}
\label{sec:quality}
Since the fuzzy edge detector based proposed telegraph total variation model is claimed to be an improvement over the existing diffusion models, our main aim is to compare the edge detection and denoising results, in terms of both visual quality and quantitative measures. Therefore, to evaluate the ability of the proposed model, quantitative comparisons in terms of PSNR\cite{gonzalez2002digital}, MSSIM\cite{wang2004image}, speckle index (SI)\cite{dewaele1990comparison} and blind/referenceless image spatial quality evaluator (BRISQUE)\cite{mittal2012no} are shown with existing models. A higher value of MSSIM and PSNR confirm that the recovered output is closer to the ground truth information. Whereas, for the optimal filtering, computed values of SI and BRISQUE should be minimum. Another typical qualitative measures is also computed in terms of the ratio image, which can be defined as the point-by-point ratio between the degraded and the despeckled image \cite{argenti2013tutorial}.

\subsection{Results on Synthetic and Natural Images}
\label{sec:natural}
\begin{figure}
\begin{center}
 \begin{subfigure}[b]{0.2\textwidth}           
                \includegraphics[scale=0.40]{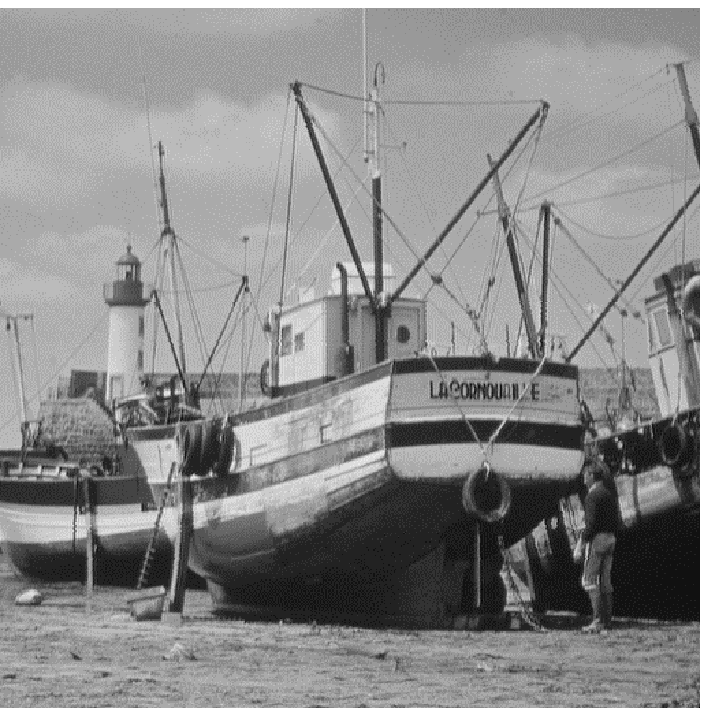}               
                \caption{Boat}
                \label{fig:circle_1_a}
        \end{subfigure} 
        \begin{subfigure}[b]{0.2\textwidth}           
                \includegraphics[scale=0.80]{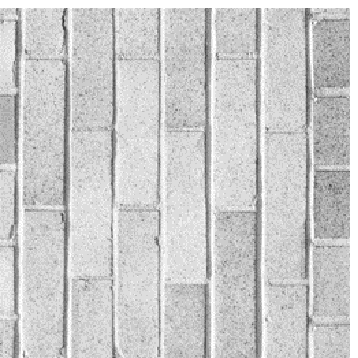}               
                \caption{Brick}
                \label{fig:circle_1_b}
        \end{subfigure}%
        
        \begin{subfigure}[b]{0.2\textwidth}
                \includegraphics[scale=0.65]{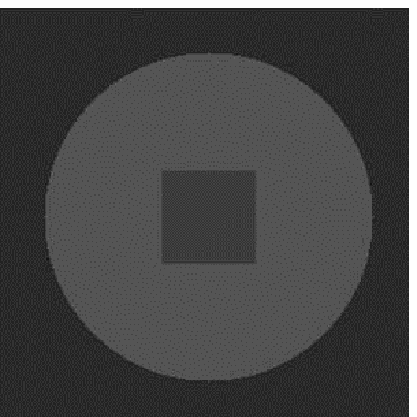}
                \caption{Circle}
                \label{fig:circle_1_c}
        \end{subfigure} 
\begin{subfigure}[b]{0.2\textwidth}
                \includegraphics[scale=0.65]{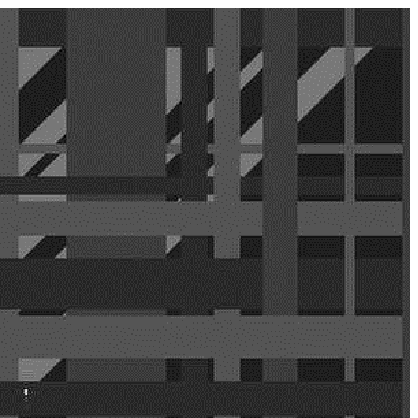}
                \caption{Texture}
                \label{fig:circle_1_d}
        \end{subfigure}  
 \begin{subfigure}[b]{0.2\textwidth}
                \includegraphics[scale=0.40]{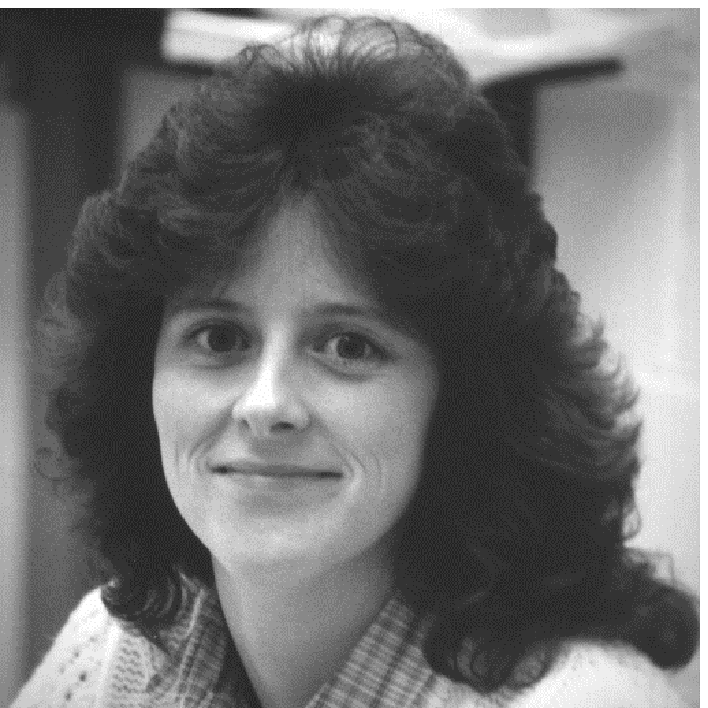}
                \caption{Woman}
                \label{fig:circle_1_e}
        \end{subfigure}          
          
\caption{ Test Images.}\label{fig:all_images}
\end{center}  
\end{figure}

To judge the despeckling ability of the proposed approach, various experiments are carried out using different synthetic as well as natural grayscale images (see figure \ref{fig:all_images} ) which are degraded by speckle noise with the different looks ($L=\lbrace1,3,5,10,33\rbrace$).

\begin{figure}
\begin{center}

        \begin{subfigure}[b]{0.2\textwidth}           
                   \includegraphics[scale=0.4]{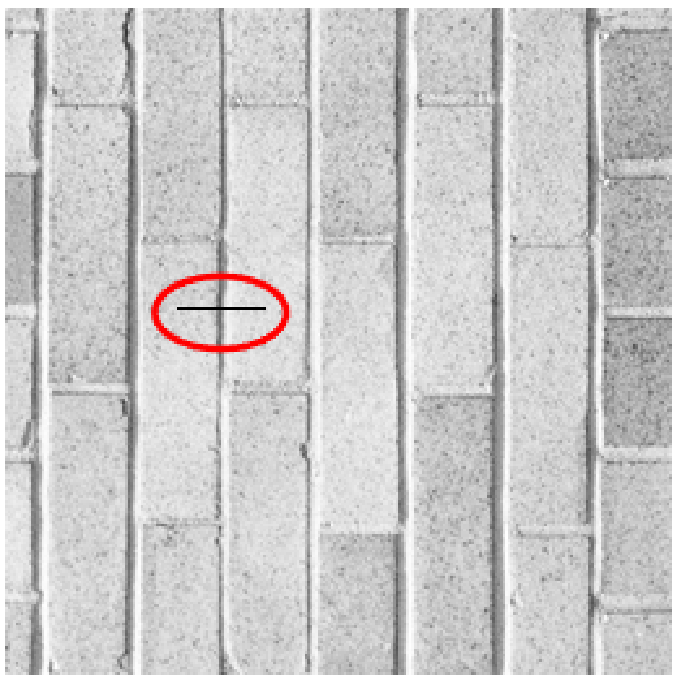}          
                \caption{}
                \label{fig1:brick_1_a}
        \end{subfigure} 
        \begin{subfigure}[b]{0.2\textwidth}           
                \includegraphics[scale=0.3]{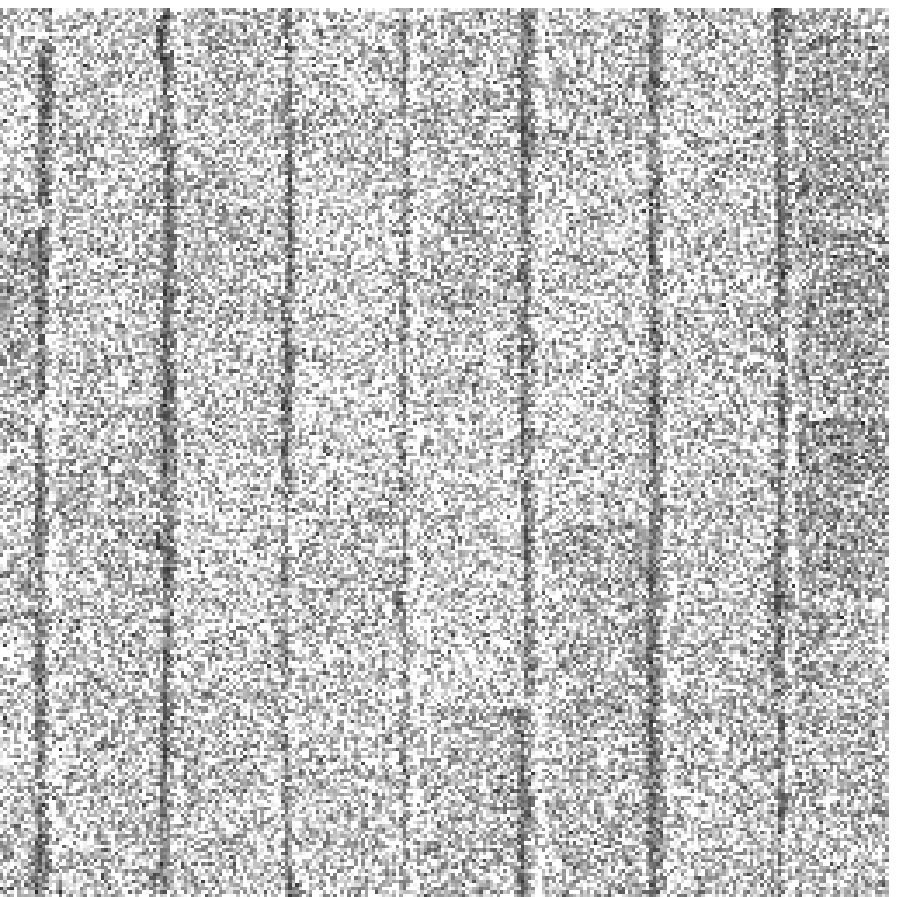}               
                \caption{}
                \label{fig1:brick_1_b}
        \end{subfigure}
         \begin{subfigure}[b]{0.2\textwidth}
                \includegraphics[scale=0.3]{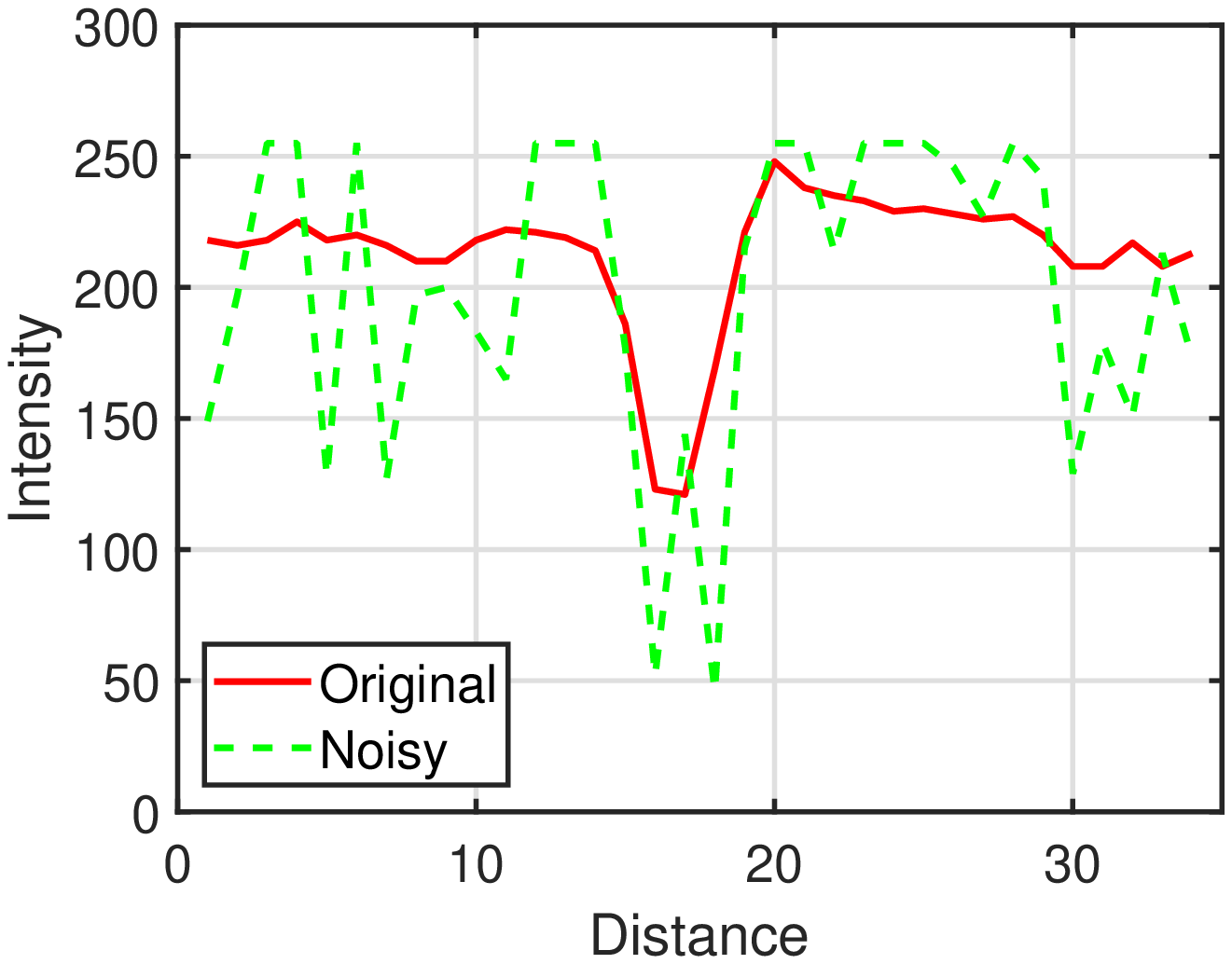}
                \caption{}
                \label{fig1:brick_1_c}
        \end{subfigure}

        \begin{subfigure}[b]{0.2\textwidth}           
                \includegraphics[scale=0.3]{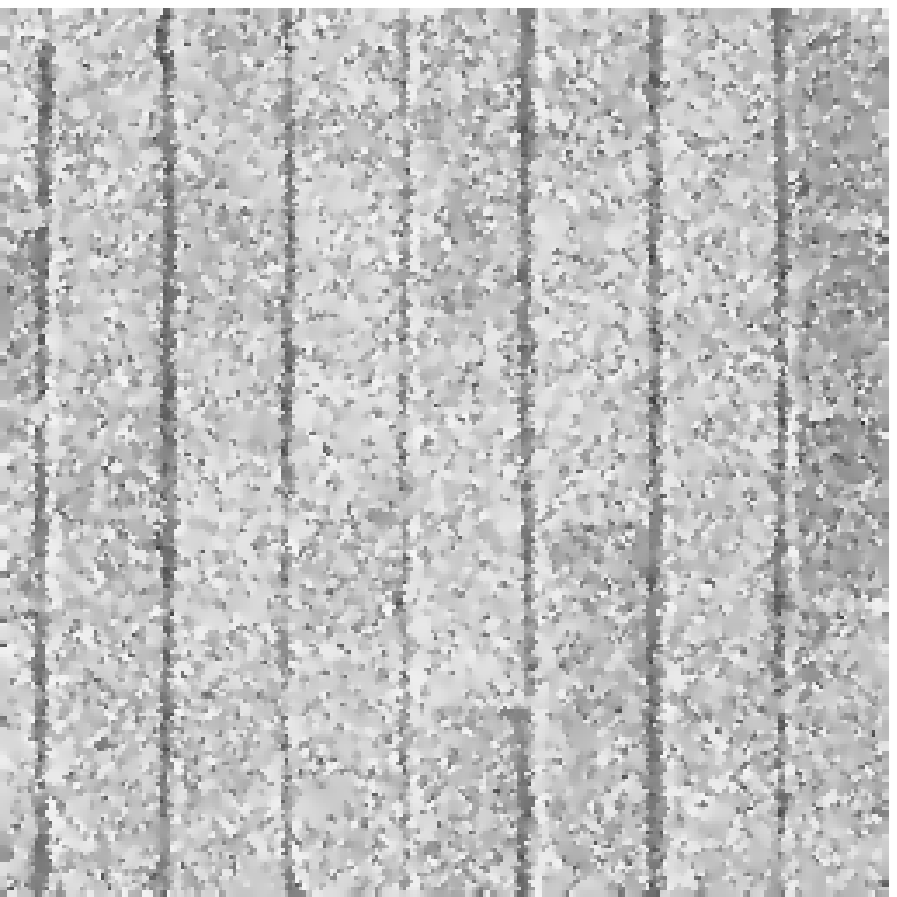}               
                \caption{}
                \label{fig1:brick_1_d}
        \end{subfigure} 
        \begin{subfigure}[b]{0.2\textwidth}           
                \includegraphics[scale=0.3]{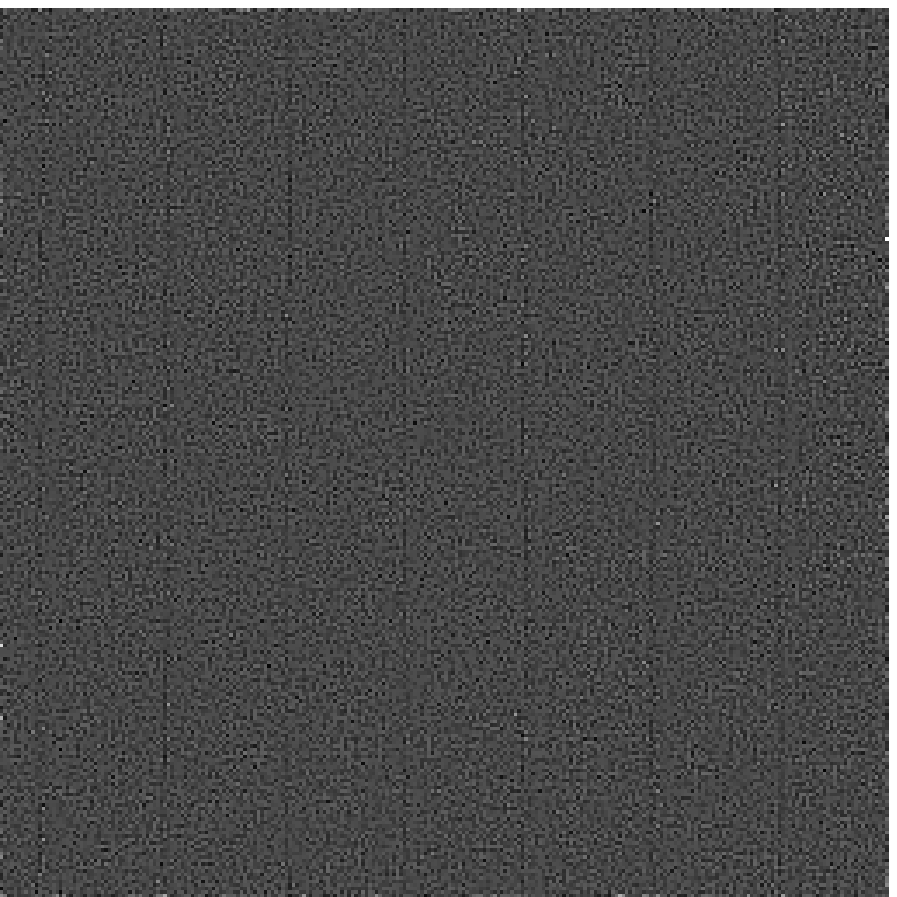}               
                \caption{}
                \label{fig1:brick_1_e}
        \end{subfigure}%
        \begin{subfigure}[b]{0.2\textwidth}
                \includegraphics[scale=0.3]{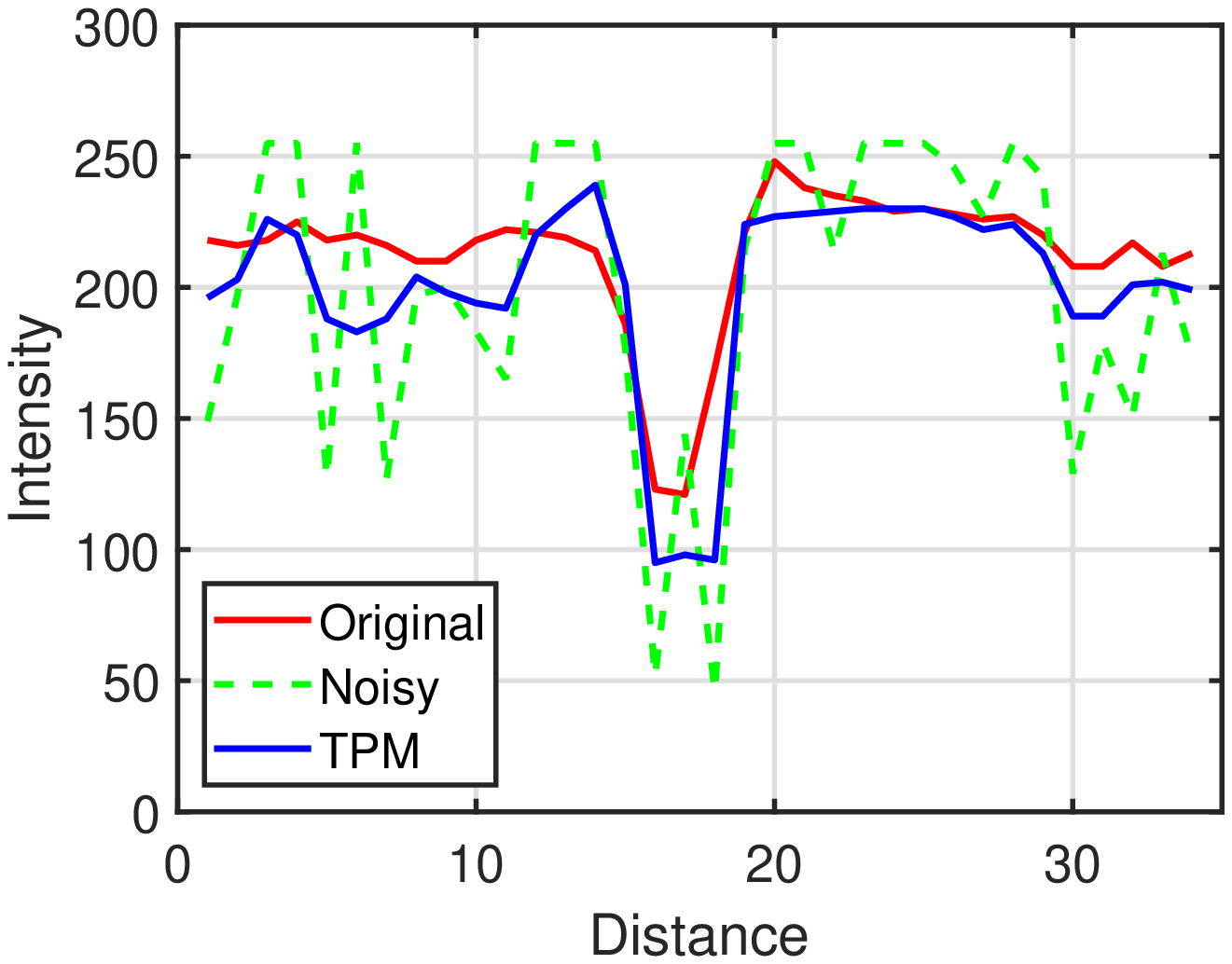}
                \caption{}
                \label{fig1:brick_1_f}
        \end{subfigure}

        \begin{subfigure}[b]{0.2\textwidth}
                \includegraphics[scale=0.3]{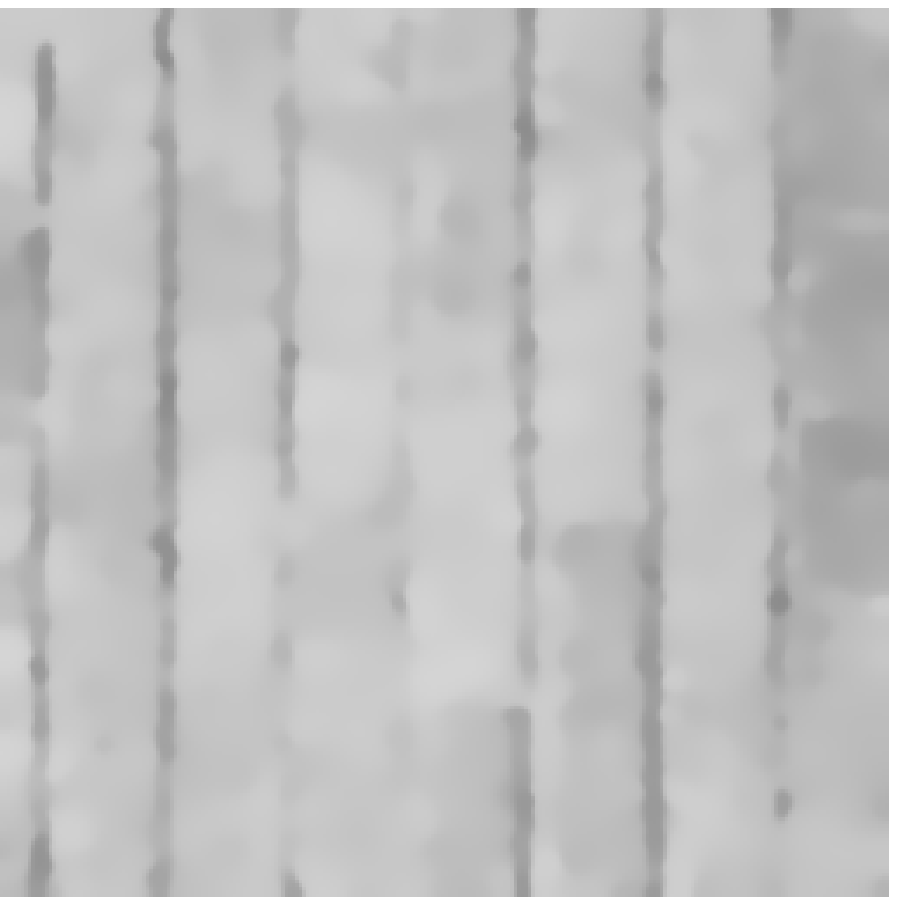}
                \caption{}
                \label{fig1:brick_1_i}
        \end{subfigure} 
        \begin{subfigure}[b]{0.2\textwidth}
                \includegraphics[scale=0.3]{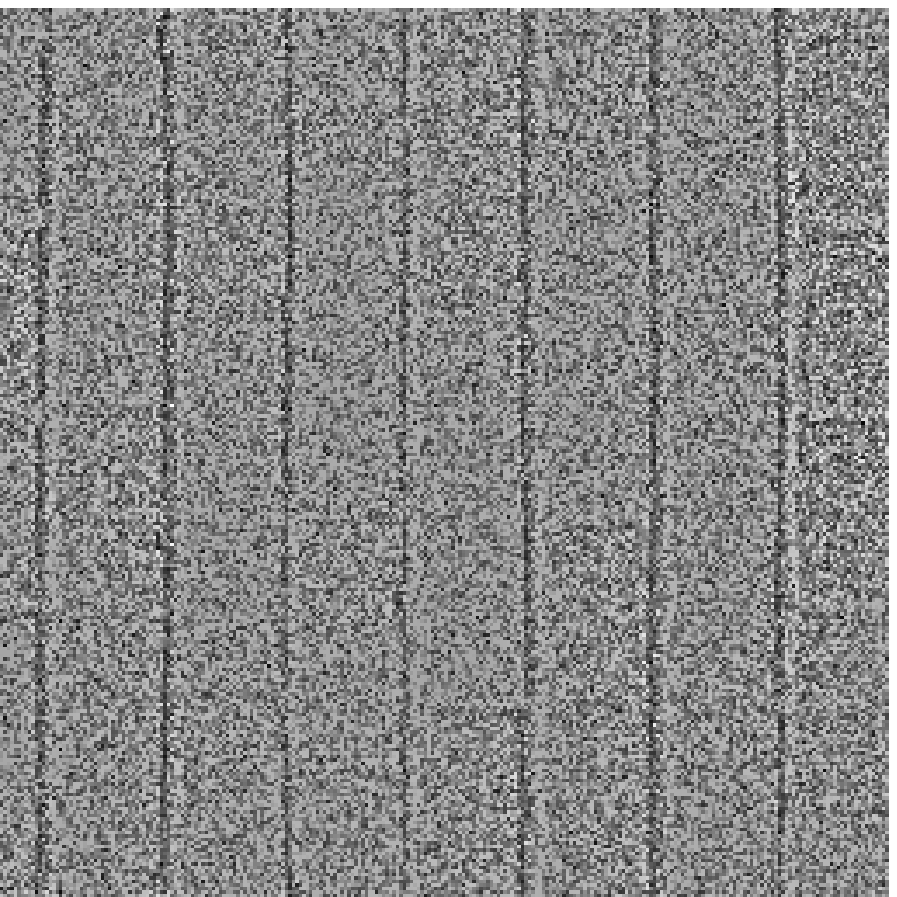}
                \caption{}
                \label{fig1:brick_1_j}
        \end{subfigure}       
        \begin{subfigure}[b]{0.2\textwidth}
                \includegraphics[scale=0.3]{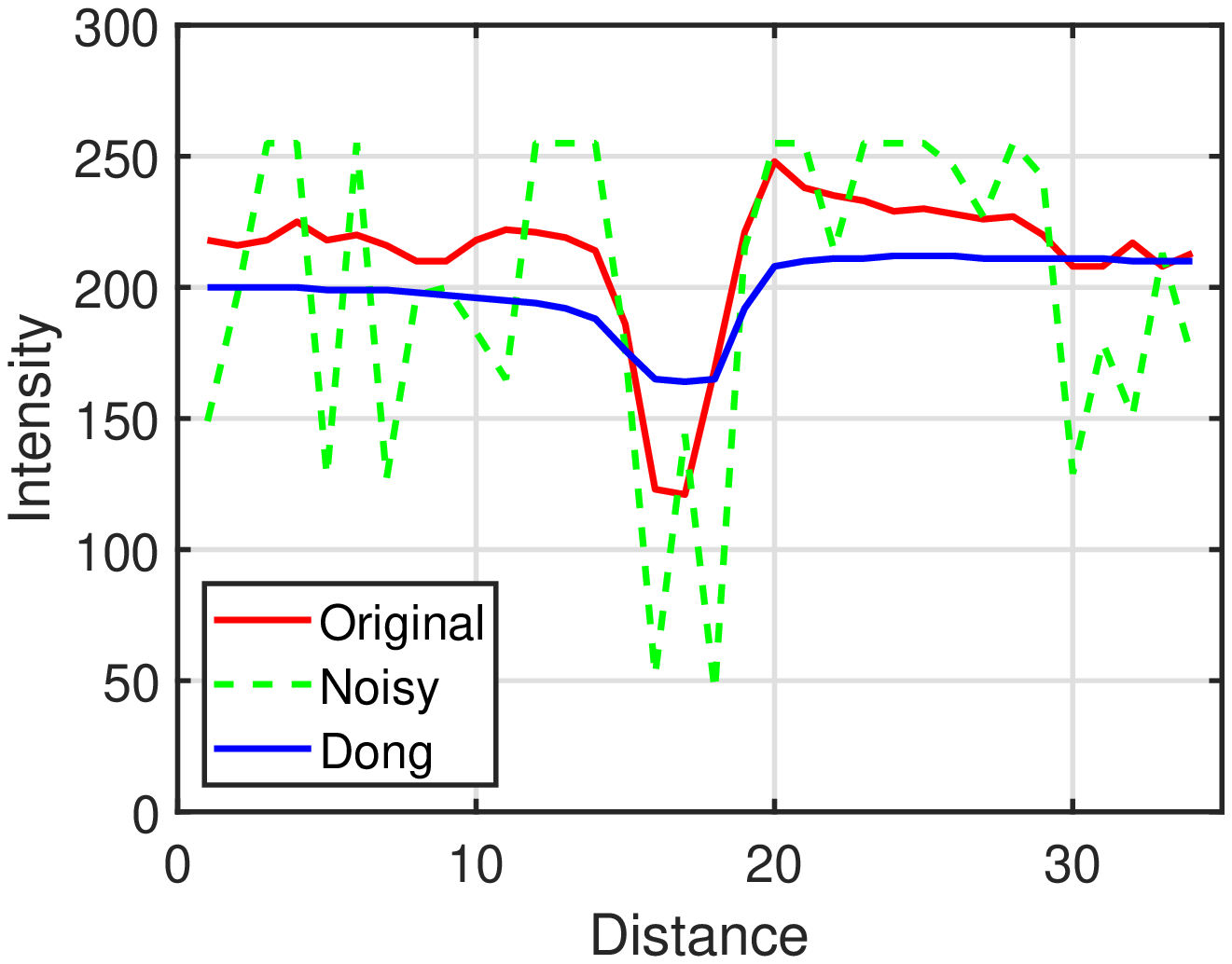}
                \caption{}
                \label{fig1:brick_1_k}
        \end{subfigure}

       \begin{subfigure}[b]{0.2\textwidth}
                \includegraphics[scale=0.75]{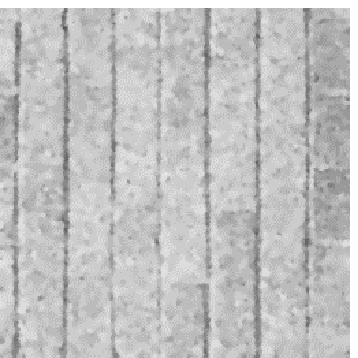}
                \caption{}
                \label{fig1:brick_1_i1}
        \end{subfigure} 
        \begin{subfigure}[b]{0.2\textwidth}
                \includegraphics[scale=0.75]{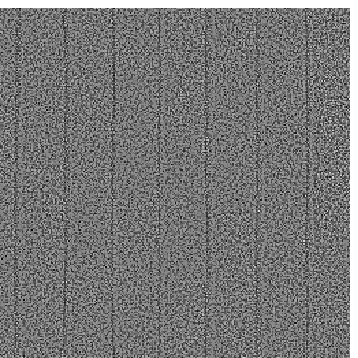}
                \caption{}
                \label{fig1:brick_1_j1}
        \end{subfigure} 
            \begin{subfigure}[b]{0.2\textwidth}
                \includegraphics[scale=0.3]{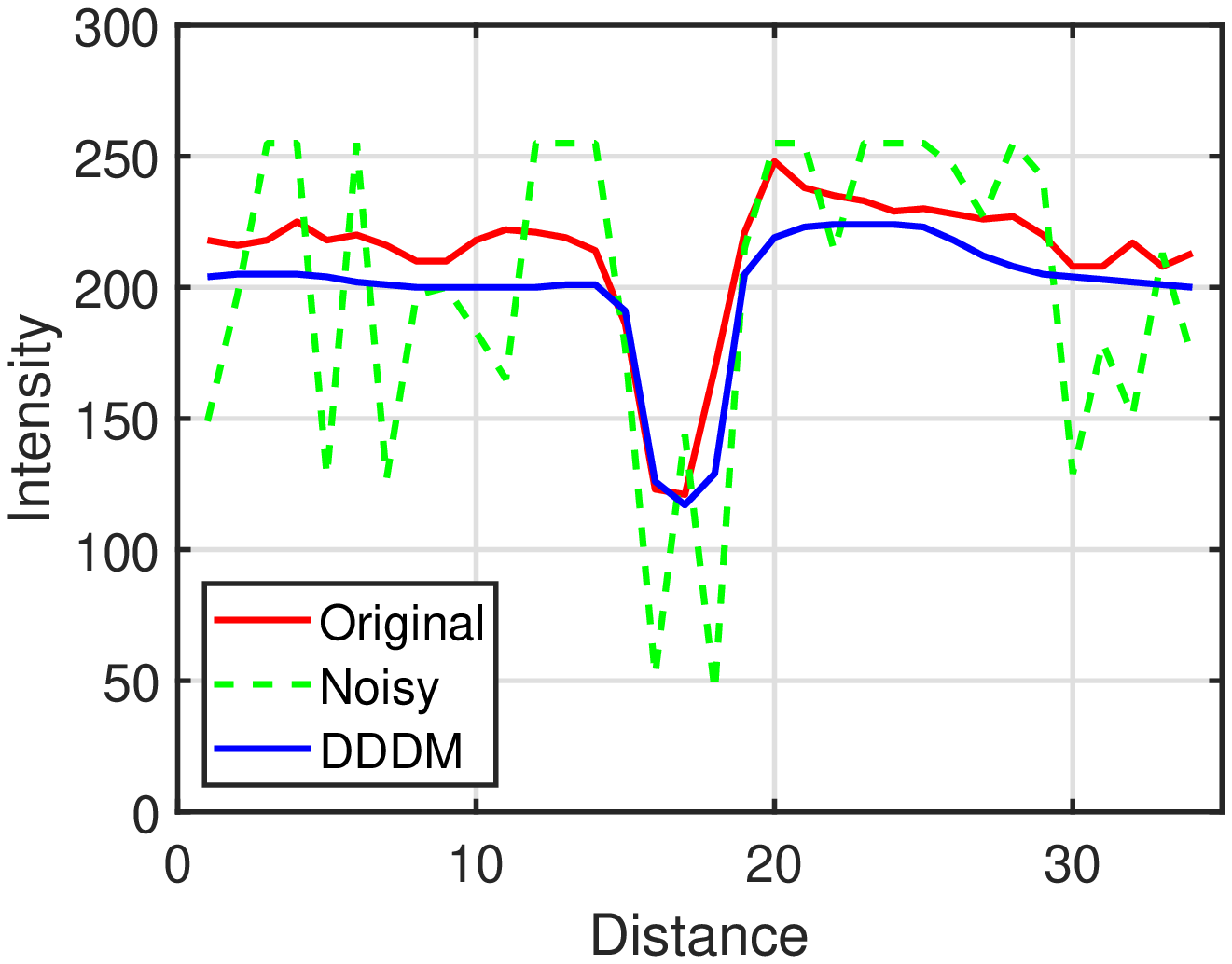}
                \caption{}
                \label{fig1:brick_1_k1}
        \end{subfigure}

  \begin{subfigure}[b]{0.2\textwidth}
                \includegraphics[scale=0.31]{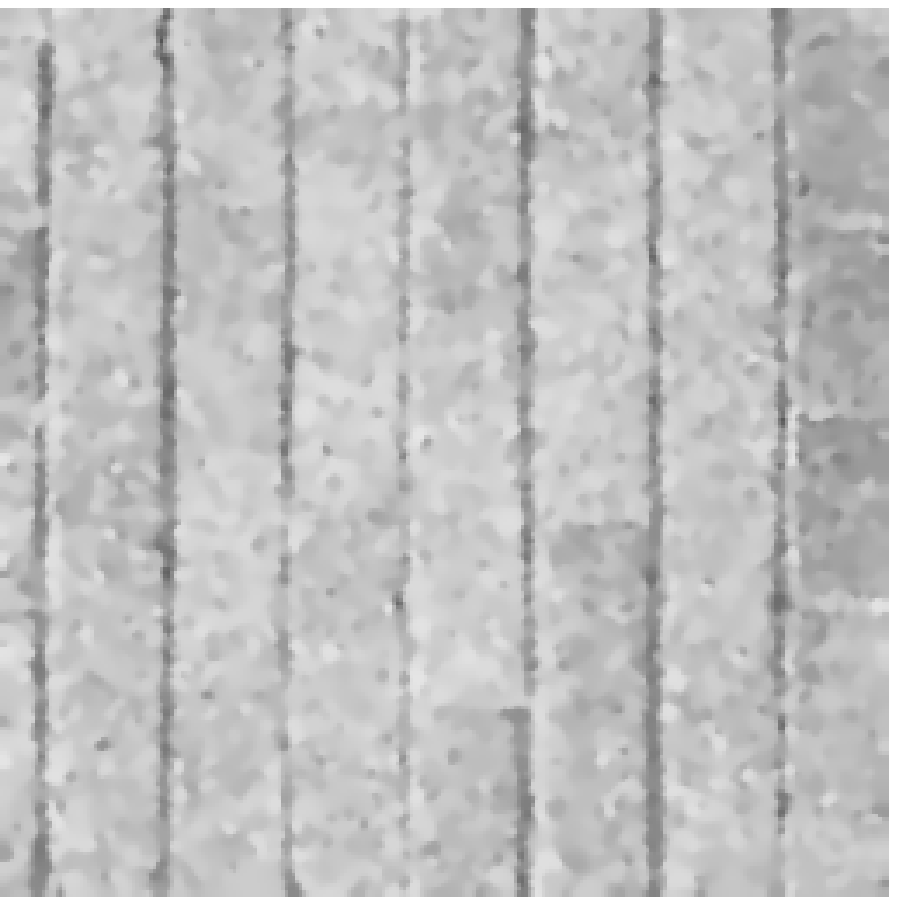}
                \caption{}
                \label{fig:brick_1_l}
        \end{subfigure}
        \begin{subfigure}[b]{0.2\textwidth}
                \includegraphics[scale=0.31]{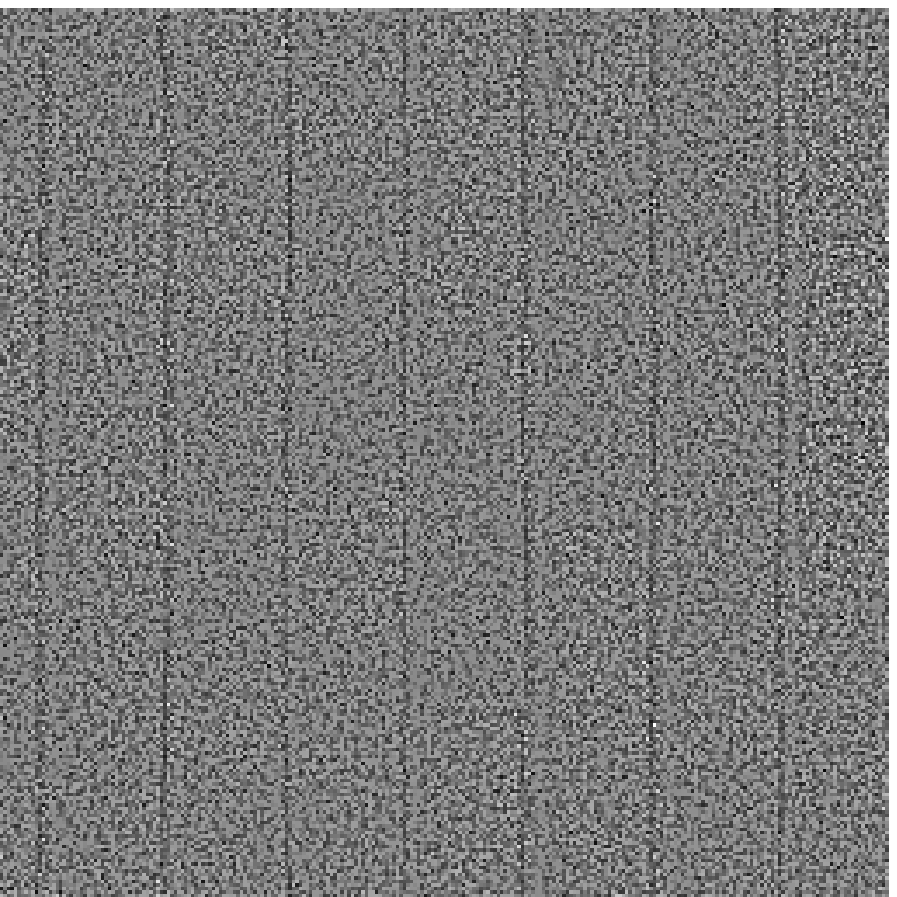}
                \caption{}
                \label{fig:brick_1_m}
        \end{subfigure}
        \begin{subfigure}[b]{0.2\textwidth}
                \includegraphics[scale=0.3]{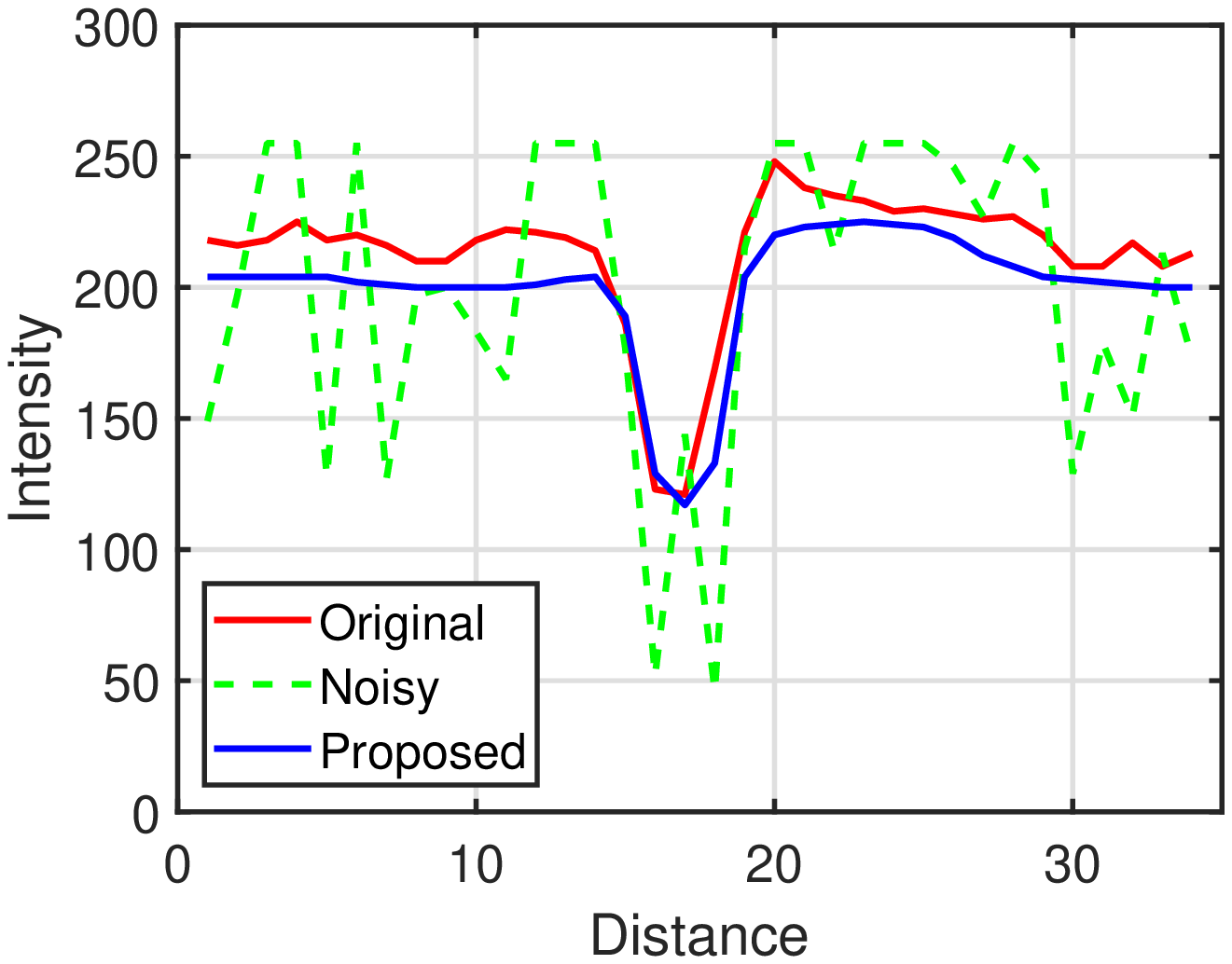}
                \caption{}
                \label{fig:brick_1_n}
        \end{subfigure}
\caption{(a) Original (b) Noisy: $L$ = 10 (c) Line profile (d-f) TPM (g-i) Dong (j-l) DDDM (m-o) Proposed. First column: Images. Middle column: Ratio images. Last column: Line profile showing 1D details. }\label{fig:Brick_L_10}
\end{center}
\end{figure}

\begin{figure}
        \centering       
        \begin{subfigure}[b]{0.3\textwidth}           
                \includegraphics[scale=0.5]{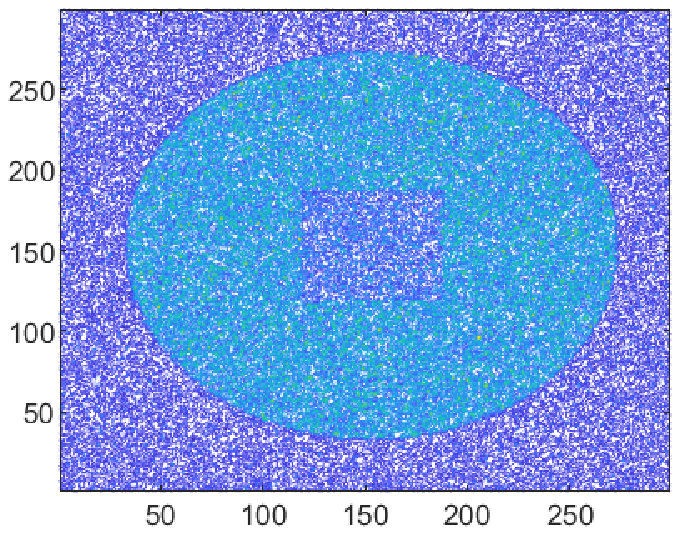}               
                \caption{}
                \label{fig:circle_detail_1_a}
        \end{subfigure} 
        \begin{subfigure}[b]{0.3\textwidth}           
                \includegraphics[scale=0.4]{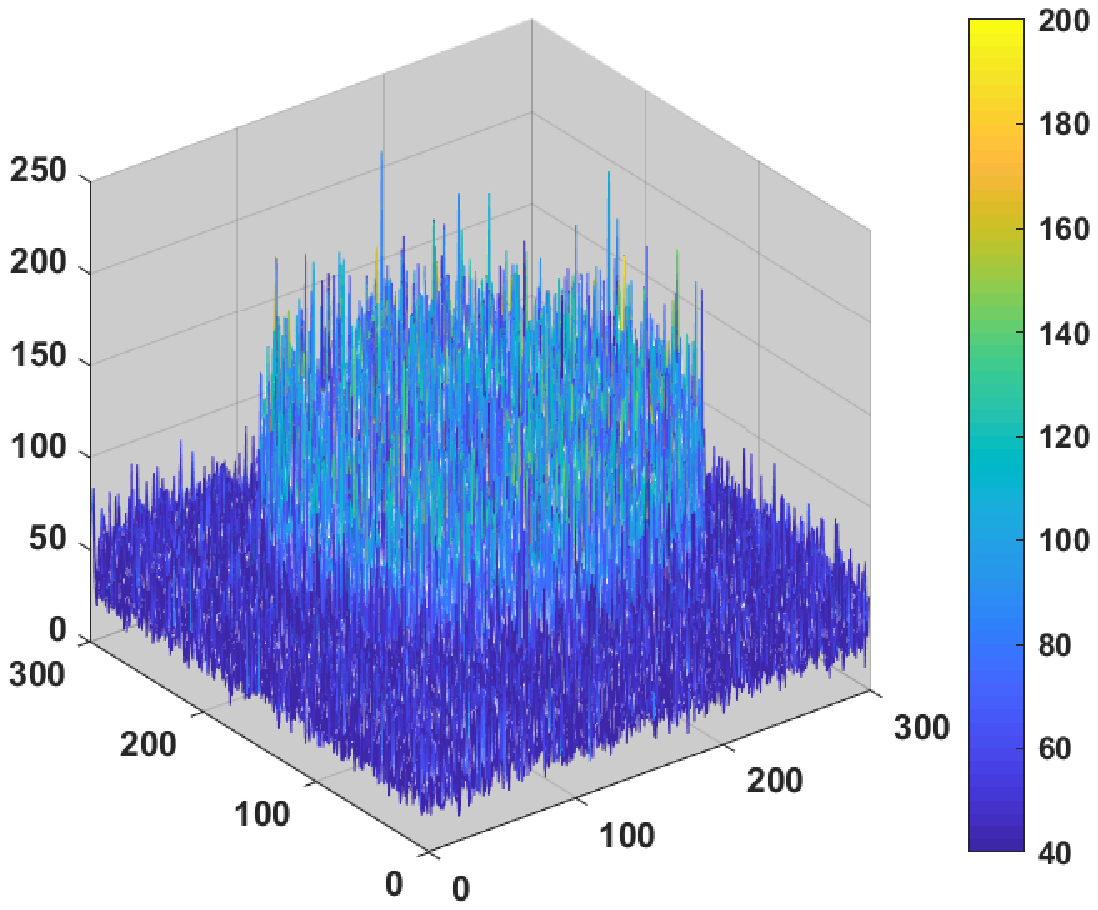}            
                \caption{}
                \label{fig:circle_detail_1_b}
        \end{subfigure}%

        \begin{subfigure}[b]{0.3\textwidth}
                \includegraphics[scale=0.3]{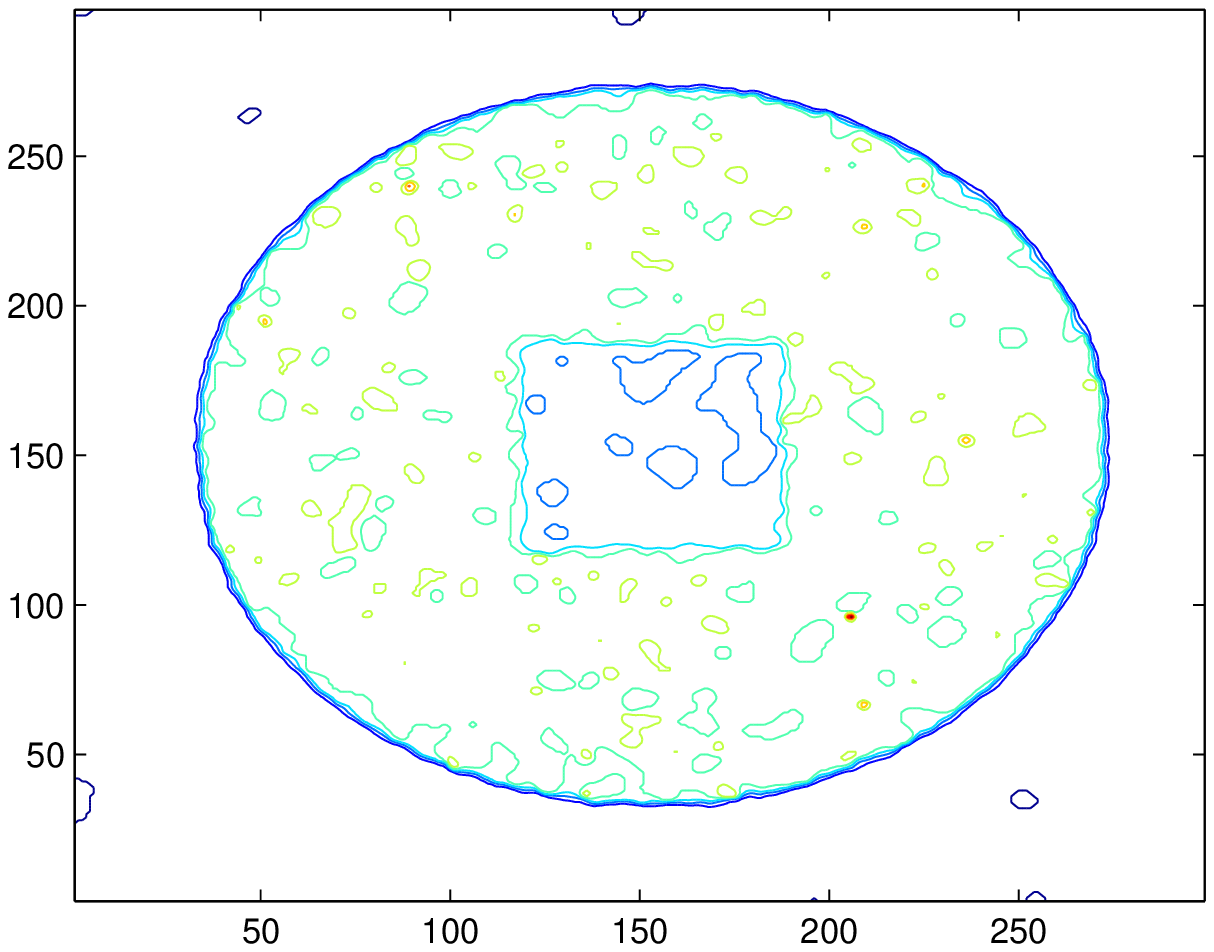}
                \caption{}
                \label{fig:circle_detail_1_g}
        \end{subfigure} 
        \begin{subfigure}[b]{0.3\textwidth}
                 \includegraphics[scale=0.4]{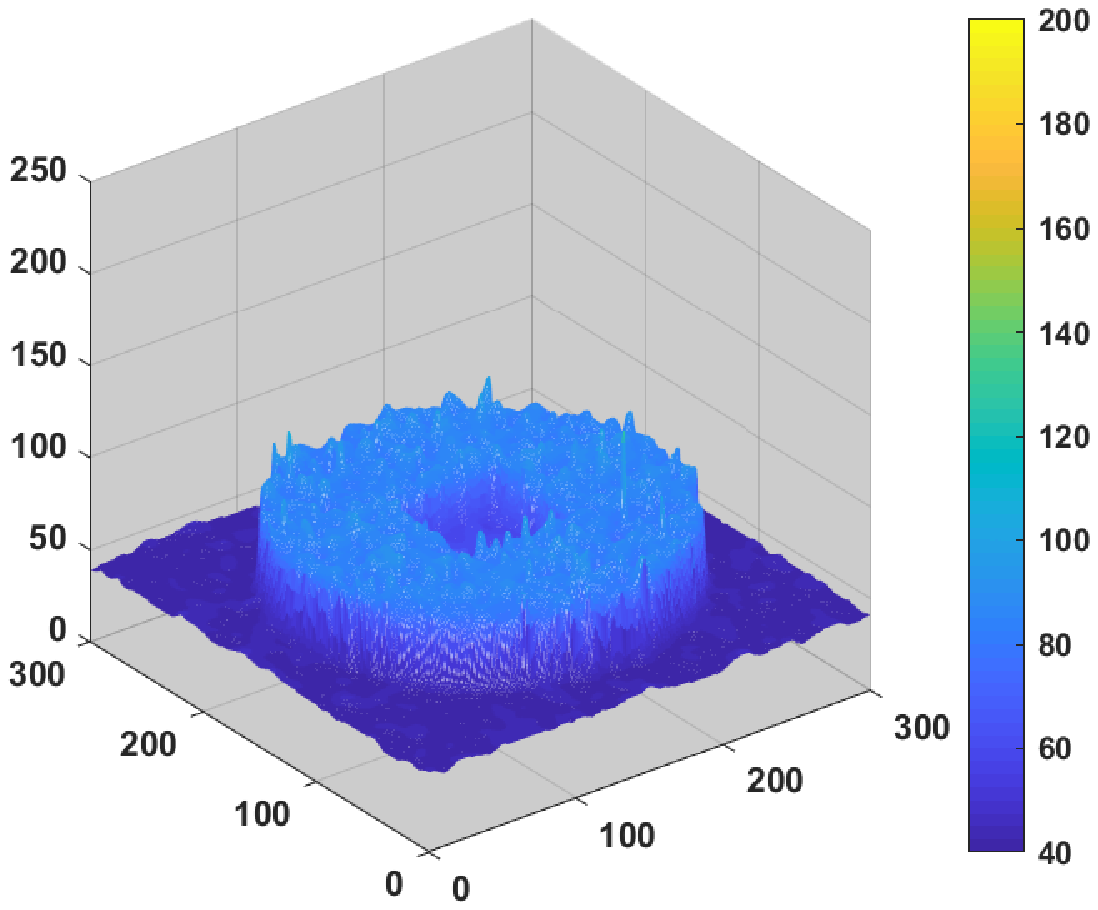}
                \caption{}
                \label{fig:circle_detail_1_h}
        \end{subfigure} 
              
           \hspace*{-0.8cm}   
           \begin{subfigure}[b]{0.3\textwidth}
                \includegraphics[scale=0.3]{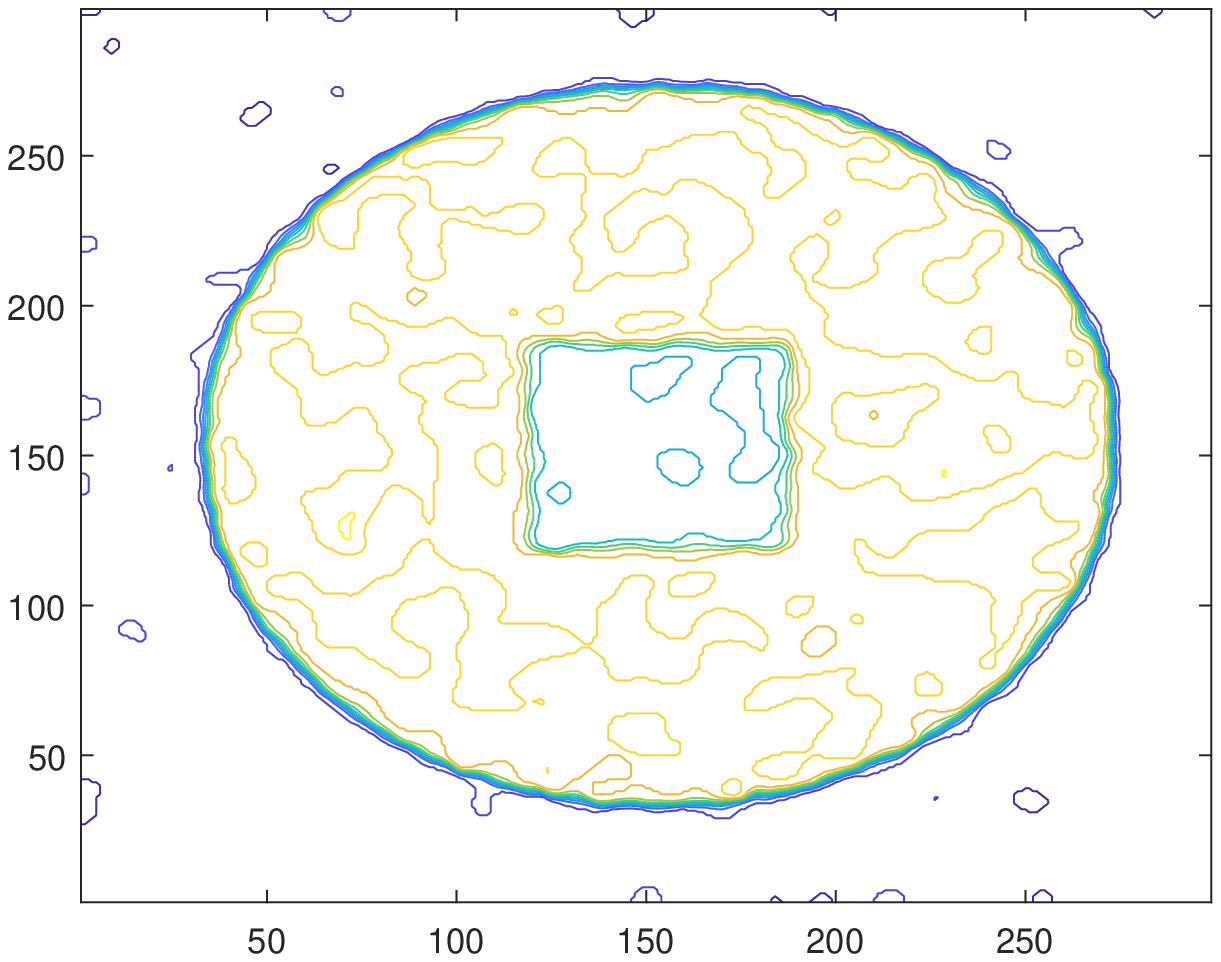}
                \caption{}
                \label{fig1:circle_detail_1_i}
        \end{subfigure} 
        \begin{subfigure}[b]{0.3\textwidth}
                     \includegraphics[scale=0.4]{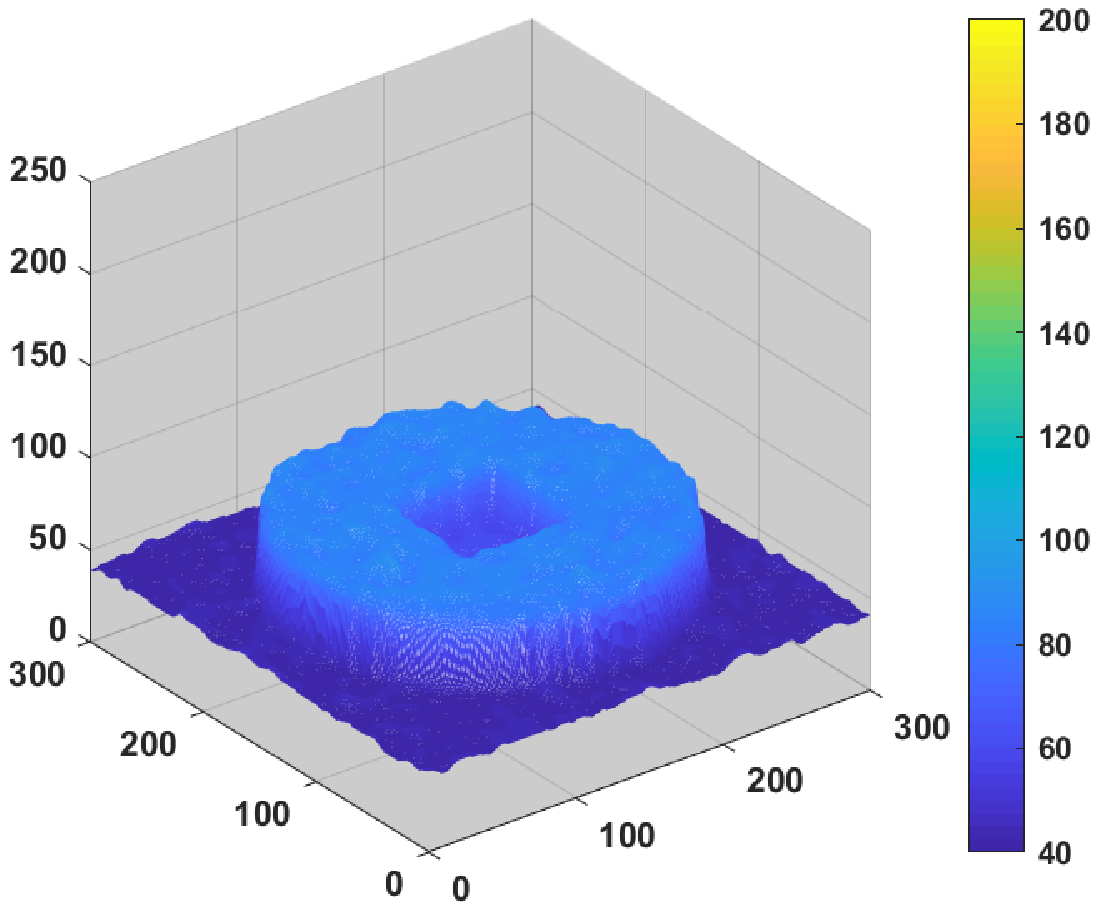}
                \caption{}
                \label{fig1:circle_detail_1_j}
        \end{subfigure}

    \begin{subfigure}[b]{0.3\textwidth}
                \includegraphics[scale=0.3]{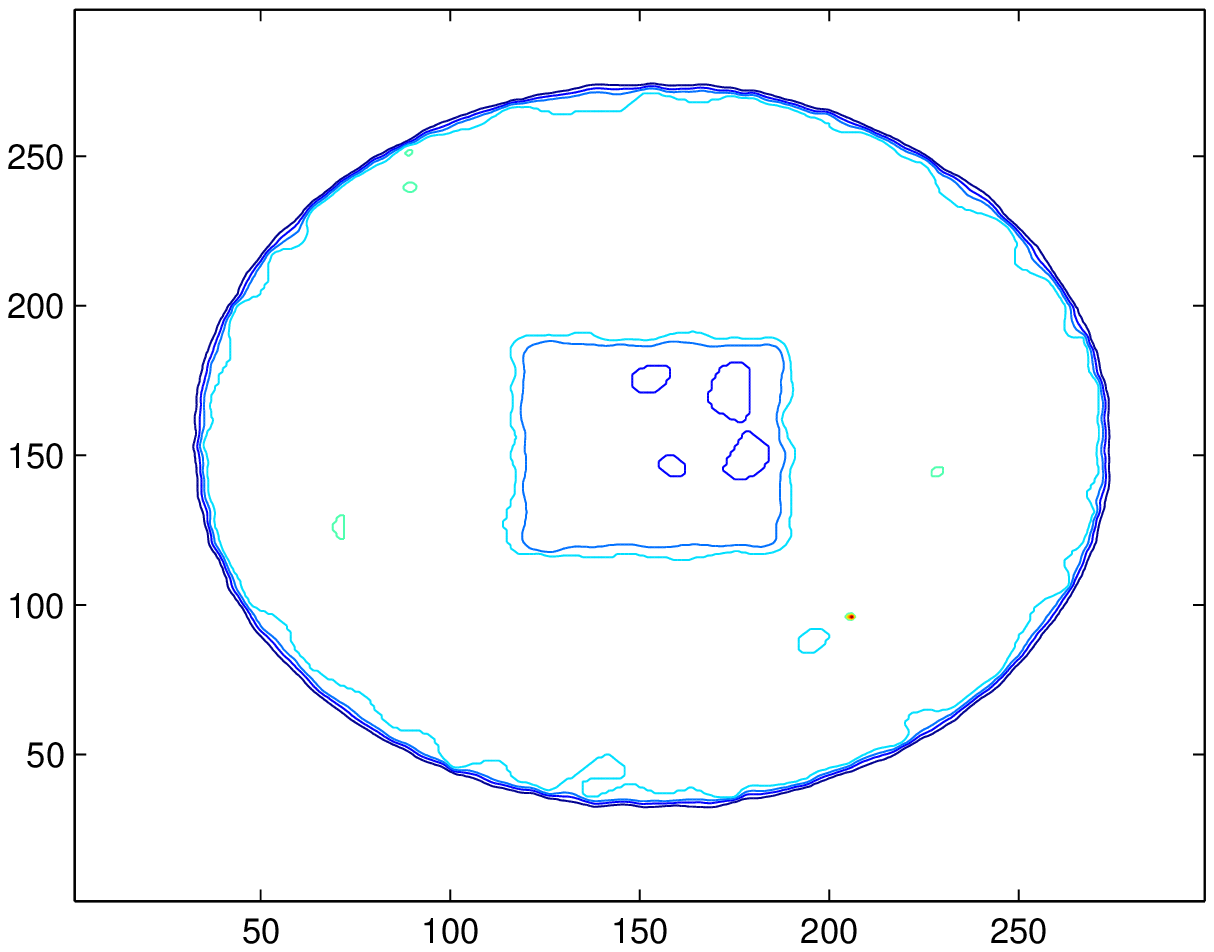}
                \caption{}
                \label{fig:circle_detail_1_i1}
        \end{subfigure}      
        \begin{subfigure}[b]{0.3\textwidth}
                \includegraphics[scale=0.4]{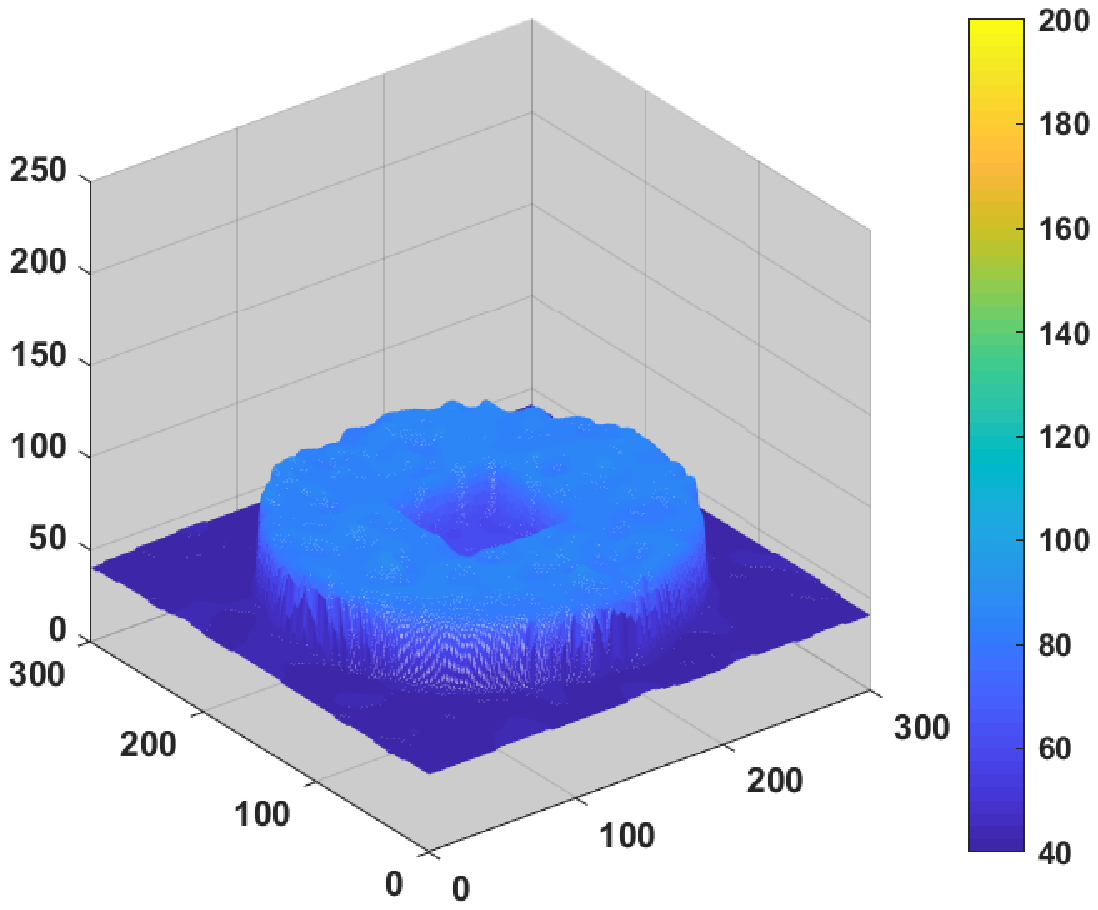}
                \caption{}
                \label{fig:circle_detail_1_j1}
        \end{subfigure}
      
\caption{Comparison of contours and 3D Surface Plots. (a-b) Noisy (c-d) Dong (e-f) DDDM (g-h) Proposed. Left column: Contours. Right column: 3D Surface.}\label{fig:Circle_Contour_10}

\end{figure}

To demonstrate the ability of the proposed method we start with a Brick image, degraded by speckle noise of look $L=10$, are displayed in figure \ref{fig:Brick_L_10}. This image contains a lot of fine texture information along with homogeneous regions. From the quality of despeckled images, it is easy to say that the restored output obtained from the TPM model is not favourable, and the texture information of the image has been degraded. Other models have also stretch texture contents of the image which results in blurred edges. To analyse the image qualitatively, we have also shown the ratio images for the brick image. By watching at the ratio image reported in the middle column of figure \ref{fig:Brick_L_10}, it is easy to observed  that the present model yields less structure content in ratio image than the other methods. Along with the despeckling of full image surface, we have also explored the same using a particular slice of the image. In this regard, the last column of figure \ref{fig:Brick_L_10}  demonstrate the slice of the clear, corrupted and restored versions. From these figures, it is effortless to judge that the proposed method eliminate speckle noise, keep sharp edges and preserve the contrast better than other discussed approaches.

To further confirm the ability of the proposed model, figure \ref{fig:Circle_Contour_10} illustrates the contour maps and 3D surface plots of the restored results for synthetic Circle image corrupted by speckle noise of look $L=10$.

It can be seen from the corresponding contour maps and 3D surface plots of the Circle image (see Fig. \ref{fig:Circle_Contour_10}), Dong and DDDM models left some serious speckles in homogeneous regions, but proposed model produces fewer artifacts with better edge preservation.

Also, the quantitative results in terms of PSNR and MSSIM values for various test images as well as for different noise levels are shown in table \ref{tab:table1}. To make the comparison convenient, the higher values for both MSSIM and PSNR  are highlighted in the table. The higher values of both measures clearly justify the superiority of the proposed approach. Further, we have demonstrated the numerical comparison with SI values of proposed and the alternative approaches, in table \ref{tab:table2}. As expected, the proposed model has the lowest SI values for different variances of the noise. Finally, after considering overall quantitative analysis (in terms of PSNR, MSSIM and SI values), we note that the proposed approach is robust and efficient in noise removal and edge preservation for the considered natural and synthetic images, compared to other models. Therefore, the telegraph total variation based despeckling framework with a fuzzy edge indicator can be recommended for most common speckle suppression tasks.

\begin{table}[h!]
\begin{center}
\caption{Comparison of MSSIM and PSNR values of despeckled images using various approaches for images corrupted by a multiplicative speckle noise with different numbers of looks (L)}
\label{tab:table1}
\scalebox{0.7}{
\begin{tabular}{ccrrrrrrrrrrrrrr}
\toprule
    \multirow{2}[8]{*}{Image} & \multirow{2}[8]{*}{L} & \multicolumn{2}{c}{TDM\cite{ratner2007image}} & \multicolumn{2}{c}{AA \cite{aubert2008variational}} & \multicolumn{2}{c}{Dong\cite{dong2013convex}} & \multicolumn{2}{c}{DDD\cite{zhou2015doubly}} & \multicolumn{2}{c}{ZZDB\cite{zhou2018nonlinear}} &   \multicolumn{2}{c}{Proposed}  \\
\cmidrule(r){3-4}
		\cmidrule(r){5-6}
		\cmidrule(r){7-8}
		\cmidrule(r){9-10}  
		\cmidrule(r){11-12} 
		\cmidrule(r){13-14} 
		 \cmidrule(r){15-16} 
		        &       & \multicolumn{1}{c}{MSSIM} & \multicolumn{1}{c}{PSNR} & \multicolumn{1}{c}{MSSIM} & \multicolumn{1}{c}{PSNR} & \multicolumn{1}{c}{MSSIM} & \multicolumn{1}{c}{PSNR} & \multicolumn{1}{c}{MSSIM}  & \multicolumn{1}{c}{PSNR} & \multicolumn{1}{c}{MSSIM} & \multicolumn{1}{c}{PSNR} & \multicolumn{1}{c}{MSSIM} & \multicolumn{1}{c}{PSNR} \\
    \midrule
Boat    & 1           & 0.3053 & 12.44 & 0.4201 & 15.90 & 0.4526 & 16.78 &0.4873 & 16.65 &0.5656 &16.98 &\textbf{0.5720} & \textbf{17.03}  \\
        & 3           & 0.4317 & 15.79 & 0.5568 & 21.14 & 0.5609 & 21.30 &0.5669 & 21.46 &0.6752 &22.30 &\textbf{0.6834} & \textbf{22.48}  \\
        & 5           & 0.4933 & 17.57 & 0.6061 & 22.76 & 0.6093 & 22.88 &0.6110 & 23.11 &0.7200 &24.14 &\textbf{0.7259} & \textbf{24.38}  \\
        & 10          & 0.5872 & 20.66 & 0.6831 & 24.91 & 0.6848 & 24.97 &0.6926 & 25.28 &0.7750 &26.09 &\textbf{0.7781} & \textbf{26.31}  \\
        & 33          & 0.7721 & 26.25 & 0.8139 & 28.16 & 0.8155 & 28.19 &0.8192 & 28.53 &0.8465 &28.62 &\textbf{0.8523} & \textbf{28.75}  \\
        &             &        &       &        &       &        &       &                 &                \\
Brick   & 1           & 0.0841 & 9.97  & 0.2593 & 11.71 & 0.2798 & 12.18 &0.2872 &12.18 &0.2880 &12.19 & \textbf{0.2888} & \textbf{12.23} \\
        & 3           & 0.1757 & 13.92 & 0.3558 & 16.93 & 0.3505 & 16.96 &0.3612 &16.72 &0.3568 &16.78 & \textbf{0.3754} & \textbf{17.05} \\
        & 5           & 0.2417 & 16.04 & 0.3718 & 18.95 & 0.3650 & 18.93 &0.3936 &18.51 &0.3909 &18.54 & \textbf{0.4196} & \textbf{19.28} \\
        & 10          & 0.3594 & 19.27 & 0.3845 & 20.92 & 0.3772 & 20.91 &0.4796 &20.63 &0.4576 &21.14 & \textbf{0.4880} & \textbf{21.96}  \\
        & 33          & 0.4872 & 24.00 & 0.3985 & 22.31 & 0.3896 & 22.14 &0.4863 &21.82 &0.5122 &23.34 & \textbf{0.5971} & \textbf{25.25}  \\
        &             &        &       &        &       &        &       &                 &                \\
Circle  & 1           & 0.7621 & 27.13 & 0.7183 & 26.26 & 0.9057 & 32.22 &0.9245 &32.40 &0.9430 &33.48 & \textbf{0.9544} &\textbf{33.86}  \\
        & 3           & 0.8524 & 30.99 & 0.9271 & 34.98 & 0.9373 & 35.46 &0.9417 &35.85 &0.9603 &36.71 & \textbf{0.9656} &\textbf{36.97}  \\
        & 5           & 0.9175 & 34.47 & 0.9397 & 36.24 & 0.9447 & 36.47 &0.9521 &36.94 &0.9634 &37.58 & \textbf{0.9684} &\textbf{37.91}  \\
        & 10          & 0.9464 & 37.47 & 0.9576 & 38.46 & 0.9608 & 38.62 &0.9623 &39.08 &0.9732 &39.49 & \textbf{0.9751} &\textbf{39.66}  \\
        & 33          & 0.9741 & 41.25 & 0.9755 & 41.48 & 0.9761 & 41.52 &0.9781 &41.58 &0.9791 &41.72 & \textbf{0.9820} &\textbf{42.02}  \\
        &             &        &       &        &       &        &       &                 &                \\
Texture & 1           & 0.7073 & 24.71 & 0.6840 & 24.20 & 0.8040 & 26.44 & 0.7553 & 26.82 & 0.7580 & 26.97 & \textbf{0.8050} & \textbf{27.14} \\
        & 3           & 0.8053 & 27.33 & 0.8574 & 29.89 & 0.8227 & 27.69 & 0.8006 & 29.22 & 0.8285 & 29.77 & \textbf{0.8700} & \textbf{29.95} \\
        & 5           & 0.8162 & 27.75 & 0.8770 & 30.67 & 0.8319 & 28.10 & 0.8205 & 30.11 & 0.8458 & 30.67 & \textbf{0.8851} & \textbf{30.88} \\
        & 10          & 0.8178 & 27.84 & 0.8996 & 31.70 & 0.8323 & 28.20 & 0.8503 & 31.71 & 0.8570 & 31.73 & \textbf{0.9014} & \textbf{32.44} \\
        & 33          & 0.8198 & 27.87 & 0.9048 & 32.11 & 0.8368 & 28.43 & 0.8973 & 34.25 & 0.9049 & 34.51 & \textbf{0.9263} & \textbf{34.65} \\
        &             &        &       &        &       &        &       &                 &                \\
Woman   & 1           & 0.3255 & 14.22 & 0.5024 & 16.96 & 0.6309 & 17.71 & 0.7086 & 17.86 & 0.7150 & 17.99 & \textbf{0.7784} & \textbf{18.05} \\
        & 3           & 0.5639 & 18.14 & 0.7158 & 22.56 & 0.7244 & 22.70 & 0.8056 & 23.41 & 0.8003 & 23.52 & \textbf{0.8510} & \textbf{23.77} \\
        & 5           & 0.6487 & 20.18 & 0.7559 & 24.41 & 0.7610 & 24.53 & 0.8411 & 25.64 & 0.8298 & 25.72 & \textbf{0.8719} & \textbf{26.10} \\
        & 10          & 0.7483 & 23.57 & 0.8126 & 27.20 & 0.8146 & 27.28 & 0.8792 & 28.33 & 0.8714 & 28.91 & \textbf{0.8957} & \textbf{29.35} \\
        & 33          & 0.8772 & 29.73 & 0.8953 & 31.46 & 0.8954 & 31.47 & 0.9310 & 33.04 & 0.9267 & 33.18 & \textbf{0.9321} & \textbf{33.50} \\
\bottomrule
\end{tabular}
}
\end{center}
\end{table}
\begin{table}[h!]
\centering
\caption{Comparison of Speckle index of despeckled images}
\label{tab:table2}
\scalebox{0.7}{
\begin{tabular}{ccrrrrrr}
\toprule
Image  & Noise Level (L) & TDM\cite{ratner2007image} & AA\cite{aubert2008variational}  & Dong\cite{dong2013convex} & DDD\cite{zhou2015doubly} &  ZZDB\cite{zhou2018nonlinear} & Proposed          \\
 \midrule
boat   & 1           & 0.5736 & 0.3895 & 0.3368 & 0.3417  & 0.3289 & \textbf{0.3173} \\
       & 3           & 0.4735 & 0.3759 & 0.3712 & 0.3720  & 0.3569 & \textbf{0.3472} \\
       & 5           & 0.4421 & 0.3783 & 0.3755 & 0.3762  & 0.3637 & \textbf{0.3558} \\
       & 10          & 0.4099 & 0.3796 & 0.3782 & 0.3794  & 0.3709 & \textbf{0.3658} \\
       & 33          & 0.3874 & 0.3800 & 0.3795 & 0.3809  & 0.3756 & \textbf{0.3746} \\
       &             &        &        &        &         &        &                \\
circle & 1           & 0.3481 & 0.3906 & 0.3165 & 0.3219  & 0.3098 & \textbf{0.3013} \\
       & 3           & 0.3368 & 0.3294 & 0.3245 & 0.3271  & 0.3215 & \textbf{0.3163} \\
       & 5           & 0.3296 & 0.3295 & 0.3271 & 0.3289  & 0.3249 & \textbf{0.3202} \\
       & 10          & 0.3295 & 0.3291 & 0.3279 & 0.3290  & 0.3253 & \textbf{0.3241} \\
       & 33          & 0.3280 & 0.3275 & 0.3272 & 0.3273  & 0.3262 & \textbf{0.3258} \\
       &             &        &        &        &                 \\
woman  & 1           & 0.6145 & 0.5046 & 0.4504 & 0.4663  & 0.4372 & \textbf{0.4276} \\
       & 3           & 0.5760 & 0.5188 & 0.5140 & 0.5204  & 0.5062 & \textbf{0.4985} \\
       & 5           & 0.5688 & 0.5351 & 0.5327 & 0.5365  & 0.5264 & \textbf{0.5195} \\
       & 10          & 0.5634 & 0.5497 & 0.5487 & 0.5507  & 0.5446 & \textbf{0.5397} \\
       & 33          & 0.5658 & 0.5639 & 0.5637 & 0.5649  & 0.5619 & \textbf{0.5595}\\
       \bottomrule
\end{tabular}
}
\end{table}
\subsection{Results on Real SAR Images}
\label{sec:real}
\begin{figure}

        \centering 
        \begin{subfigure}[b]{0.45\textwidth}           
                \includegraphics[scale=0.4]{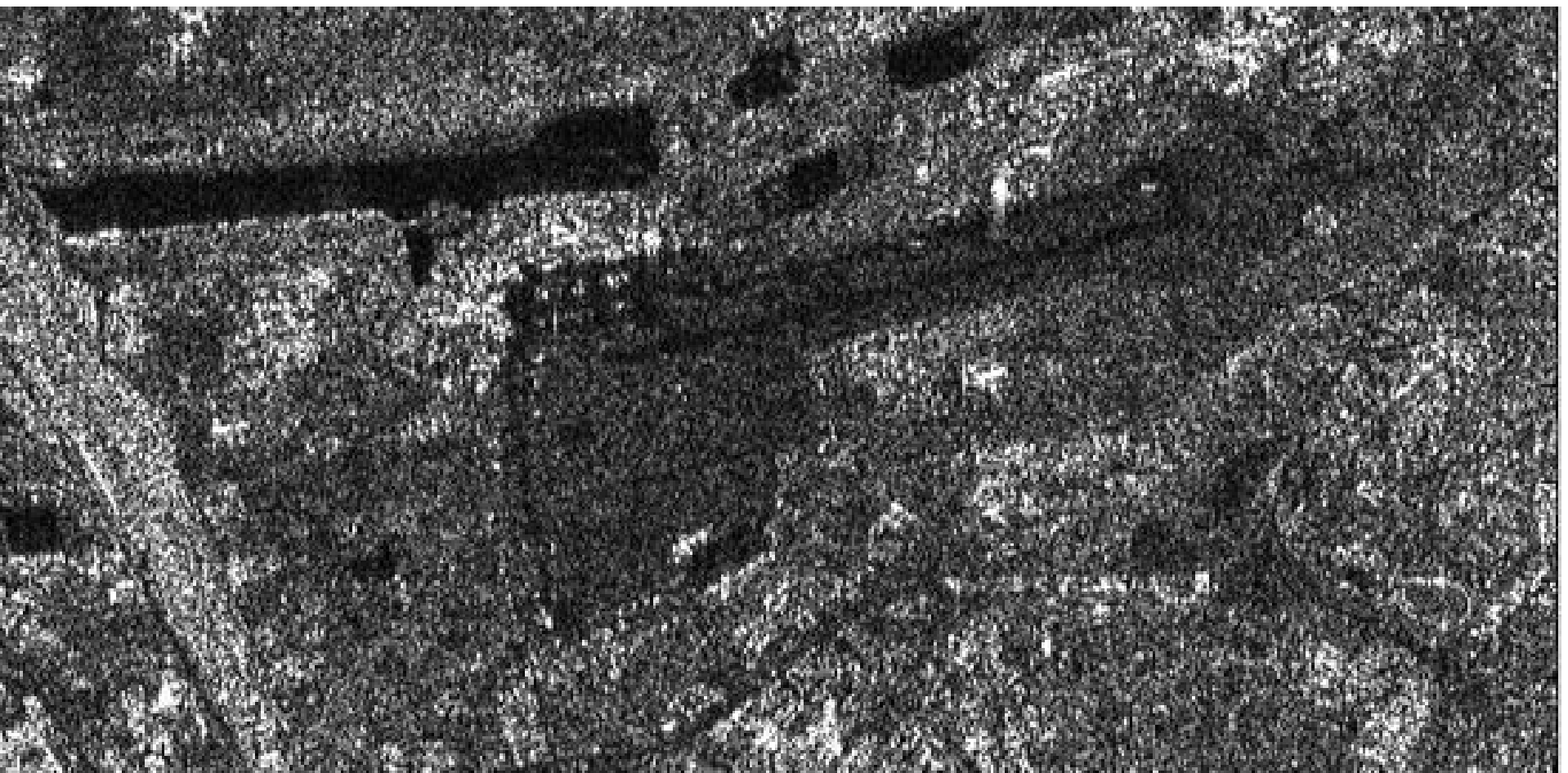}               
                \caption{}
                \label{fig:image3}
        \end{subfigure} 
         
        \begin{subfigure}[b]{0.45\textwidth}           
                \includegraphics[scale=0.5]{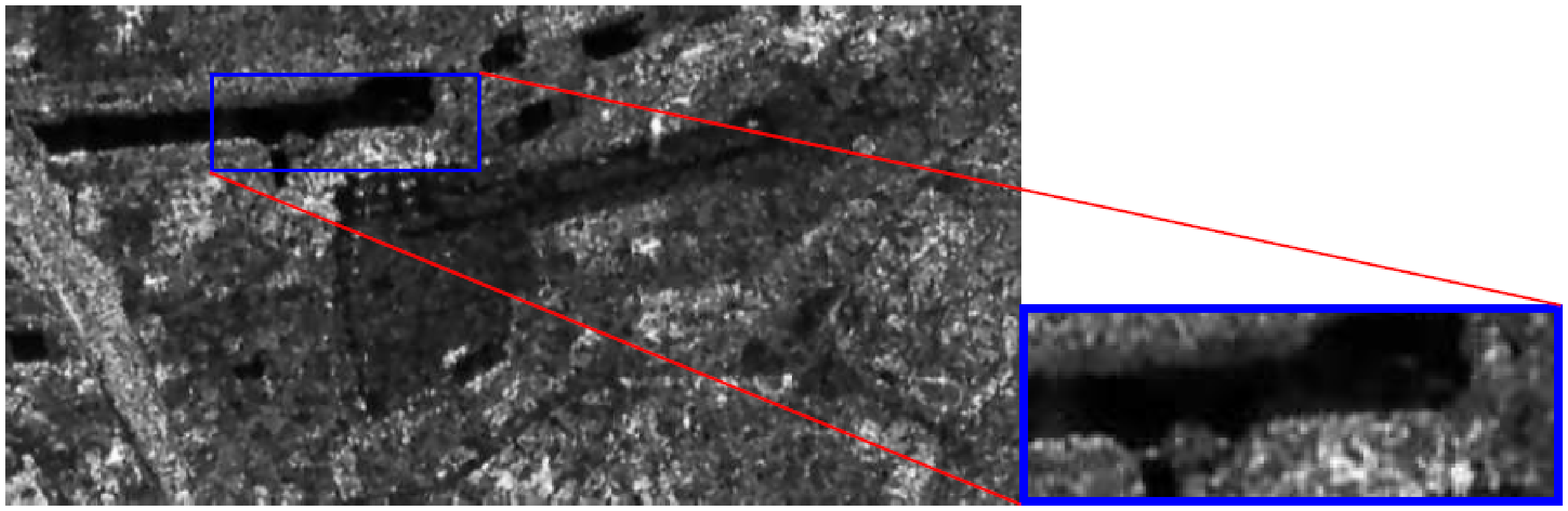}               
                \caption{}
                \label{fig:image3_new_new}
        \end{subfigure}%
        
 \caption{(a) Image1: One look radar image \cite{dataset_european}, (b) Restored image by proposed model.}\label{fig:image3_results}
\end{figure}

In figure \ref{fig:image3_results} we display the filtered image using the proposed model for a real single-look SAR image.
Observing the result in figure \ref{fig:image3_results}, one can conclude that the result seem to be well despeckled with efficient shape and edge preservation. Moreover, present model efficiently preserve the spatial resolution as well as significantly reduce the speckle effect, as can be seen from a closer look of the restoration result.

Beside the visual description of results illustrated in figure \ref{fig:image3_results}, the quantitative results of the filtered image, in term of SI index and BRISQUE value are also computed and described in table \ref{tab:table3}. Lower values of both measures for each case clearly indicates the robustness of the proposed model. Therefore, compared with existing models, it is easy to observe that the proposed method is more robust and promising for the speckle reduction problem in SAR images.
\begin{table}[]
\centering
\caption{Comparison of SI and BRISQUE(BQ) values of despeckled images using various approaches for real SAR images}
\label{tab:table3}
\scalebox{0.65}{
\begin{tabular}{ccrrrrrrrrrrrrrr}
\toprule
    \multirow{2}[4]{*}{Image} & \multirow{2}[4]{*}{Noise Level} & \multicolumn{2}{c}{TDM\cite{ratner2007image}} & \multicolumn{2}{c}{AA \cite{aubert2008variational}} & \multicolumn{2}{c}{Dong\cite{dong2013convex}} & \multicolumn{2}{c}{DDDM\cite{zhou2015doubly}} & \multicolumn{2}{c}{ZZDB\cite{zhou2018nonlinear}}  & \multicolumn{2}{c}{Proposed} \\
\cmidrule(r){3-4}
		\cmidrule(r){5-8}
		\cmidrule(r){7-8}
		\cmidrule(r){9-10}   
		\cmidrule(r){11-12} 
		 \cmidrule(r){13-14}       &       & \multicolumn{1}{c}{SI} & \multicolumn{1}{c}{BQ} & \multicolumn{1}{c}{SI} & \multicolumn{1}{c}{BQ} & \multicolumn{1}{c}{SI} & \multicolumn{1}{c}{BQ} & \multicolumn{1}{c}{SI}  &\multicolumn{1}{c}{BQ}  &\multicolumn{1}{c}{SI} & \multicolumn{1}{c}{BQ} & \multicolumn{1}{c}{SI} & \multicolumn{1}{c}{BQ} \\
    \midrule
Image1  & Single-look  & 0.5966 & 59.59 & 0.5076 & 43.21 & 0.5034 & 43.99 & 0.5283 & 42.83 & 0.4806 & 42.56 & \textbf{0.4398} & \textbf{42.45} \\
\bottomrule
\end{tabular}
}
\end{table}

\section{Conclusion and Future Scope}
\label{sec:Conclusion}
This work proposes a novel and efficient fuzzy edge detector based adaptive telegraph total variation model for speckle noise suppression. The goal of such a new adaptive filtering scheme is to preserve edges efficiently when the model is applied to image despeckling. To overcome the limitations present in the existing gradient based despeckling models, we considered a hybrid approach where we combine a robust fuzzy edge indicator function with telegraph total variation model for image selective smoothing and restoration. To the best of our knowledge, the fuzzy edge detector driven telegraph total variation model has not been used before for speckle noise suppression. The total variation in the proposed model has been fused together with the IFD function based edge indicator to define edge probabilities more accurately, which in turn provides better despeckling results.
Also, we study the well-posedness of the regularized version of the proposed model using Schauder fixed point theorem. The stopping criterion for the iterative diffusion process is computed using relative error between two successive steps. Also, to measure the denoising performance of the present model, several quantitative indexes are studied. Extensive numerical experiments have been conducted to highlight the efficiency and reliability of the proposed model for despeckling using various natural and real SAR images. From our numerical experiments, it is confirmed that the proposed model has a better ability than the existing diffusion based models to restore the images without introducing undesired artifacts. Hence the proposed telegraph total variation framework indicates a potential direction for the image denoising problem when images are degraded by speckle noise. Extending the proposed telegraph total variation framework to handle texture preservation in various real images which are losing their feature by mixed noises, is an exciting future direction, which needs to be explored further. Another important step might be the use of advanced numerical schemes to solve the proposed model to improve the convergence speed of the proposed model.

\section*{Competing interests}
The authors declare that they have no competing interests.
\thispagestyle{empty}

\bibliographystyle{unsrt}

\begin{thebibliography}{1}

\bibitem{raadams1975} R. Adam, Sobolev spaces, in: Pure and Applied Mathematics Series of
Monographs and Textbooks, Vol. 65, Academic Press, Inc., New York, San
Francisco, London,, 1975.

\bibitem{dataset_european} Agency, E.S.: Esa earth online.\url{ https://earth.esa.int/handbooks/asar/CNTR1-4.html}

\bibitem{aja2001fuzzy} Aja, S., Alberola, C., Ruiz, A.: Fuzzy anisotropic diffusion for speckle filtering. In: 2001 IEEE International Conference
on Acoustics, Speech, and Signal Processing. Proceedings (Cat. No. 01CH37221), vol. 2, pp. 1261-1264. IEEE (2001)

\bibitem{argenti2013tutorial} Argenti, F., Lapini, A., Bianchi, T., Alparone, L.: A tutorial on speckle reduction in synthetic aperture radar images.
IEEE Geoscience and remote sensing magazine 1(3), 6-35 (2013)

\bibitem{atanassov2003intuitionistic} Atanassov, K.T.: Intuitionistic fuzzy sets: past, present and future. In: EUSFLAT Conf., pp. 12-19 (2003)

\bibitem{aubert2008variational} Aubert, G., Aujol, J.F.: A variational approach to removing multiplicative noise. SIAM Journal on Applied Mathe-
matics 68(4), 925-946 (2008)

\bibitem{aubert2006mathematical} Aubert, G., Kornprobst, P.: Mathematical problems in image processing: partial differential equations and the calculus
of variations, vol. 147. Springer Science $\&$ Business Media (2006)


\bibitem{babu2016adaptive} Babu, J.J.J., Sudha, G.F.: Adaptive speckle reduction in ultrasound images using fuzzy logic on coefficient of variation.
Biomedical Signal Processing and Control 23, 93-103 (2016)

\bibitem{becerikli2005new} Becerikli, Y., Karan, T.M.: A new fuzzy approach for edge detection. In: International Work-Conference on Artificial
Neural Networks, pp. 943-951. Springer (2005)

\bibitem{binaee2014ultrasound} Binaee, K., Hasanzadeh, R.P.: An ultrasound image enhancement method using local gradient based fuzzy similarity.
Biomedical Signal Processing and Control 13, 89-101 (2014))

\bibitem{burckhardt1978speckle} Burckhardt, C.B.: Speckle in ultrasound b-mode scans. IEEE Transactions on Sonics and ultrasonics 25(1), 1-6 (1978)

\bibitem{cao2010class} Cao, Y., Yin, J., Liu, Q., Li, M.: A class of nonlinear parabolic-hyperbolic equations applied to image restoration.
Nonlinear Analysis: Real World Applications 11(1), 253-261 (2010)


\bibitem{chaira2008new} Chaira, T., Ray, A.: A new measure using intuitionistic fuzzy set theory and its application to edge detection. Applied
soft computing 8(2), 919-927 (2008)

\bibitem{chaira2003segmentation} Chaira, T., Ray, A.K.: Segmentation using fuzzy divergence. Pattern Recognition Letters 24(12), 1837-1844 (2003)

\bibitem{dewaele1990comparison} Dewaele, P., Wambacq, P., Oosterlinck, A., Marchand, J.L.: Comparison of some speckle reduction techniques for sar
images. In: Geoscience and Remote Sensing Symposium, 1990. IGARSS'90.'Remote Sensing Science for the Nineties'.,
10th Annual International, pp. 2417-2422. IEEE (1990)

\bibitem{dong2013convex} Dong, G., Guo, Z., Wu, B.: A convex adaptive total variation model based on the gray level indicator for multiplicative
noise removal. In: Abstract and Applied Analysis, vol. 2013. Hindawi Publishing Corporation (2013)

\bibitem{LCEvans1998} Evans, L.: Partial Differential Equations, in: Graduate Studies in Mathematics, vol. 19. American Mathematical
Society, Providence, Rhode Island (1998)

\bibitem{gonzalez2002digital} Gonzalez, R.C., Woods, R.E.: Digital image processing (2002)


\bibitem{ho1995fedge} Ho, K.H., Ohnishi, N.: Fedge fuzzy edge detection by fuzzy categorization and classification of edges. In: International
Workshop on Fuzzy Logic in Artificial Intelligence, pp. 182-196. Springer (1995)

\bibitem{hua2009speckle} Hua, C., Jinwen, T.: Speckle reduction of synthetic aperture radar images based on fuzzy logic. In: 2009 First
International Workshop on Education Technology and Computer Science, vol. 1, pp. 933-937. IEEE (2009)

\bibitem{jain2016edge} Jain, S.K., Ray, R.K.: Edge detectors based telegraph total variational model for image filtering. In: Information
Systems Design and Intelligent Applications, pp. 119-126. Springer (2016)

\bibitem{jain2019non} Jain, S.K., Ray, R.K.: Non-linear diffusion models for despeckling of images: achievements and future challenges. IETE
Technical Review pp. 1-17 (2019)

\bibitem{jain2015iterative} Jain, S.K., Ray, R.K., Bhavsar, A.: Iterative solvers for image denoising with diffusion models: A comparative study.
Computers $\&$ Mathematics with Applications 70(3), 191-211 (2015)

\bibitem{jain2018nonlinear} Jain, S.K., Ray, R.K., Bhavsar, A.: A nonlinear coupled diffusion system for image despeckling and application to
ultrasound images. Circuits, Systems, and Signal Processing pp. 1-30 (2018)

\bibitem{jin2000adaptive} Jin, J.S.,Wang, Y., Hiller, J.: An adaptive nonlinear diffusion algorithm for filtering medical images. IEEE Transactions
on Information Technology in Biomedicine 4(4), 298-305 (2000)

\bibitem{jin2011variational} Jin, Z., Yang, X.: A variational model to remove the multiplicative noise in ultrasound images. Journal of Mathematical
Imaging and Vision 39(1), 62-74 (2011)


\bibitem{dataset_kilauea} JPL: Space radar image of kilauea. \url{https://photojournal.jpl.nasa.gov/catalog/PIA01763}

\bibitem{klir1995fuzzy} Klir, G.J., Yuan, B.: Fuzzy sets and fuzzy logic: theory and applications, vol. 574. Prentice Hall PTR New Jersey
(1995)

\bibitem{liu2013nondivergence} Liu, Q., Li, X., Gao, T.: A nondivergence p-laplace equation in a removing multiplicative noise model. Nonlinear
Analysis: Real World Applications 14(5), 2046-2058 (2013)


\bibitem{mittal2012no} Mittal, A., Moorthy, A.K., Bovik, A.C.: No-reference image quality assessment in the spatial domain. IEEE Transac-
tions on Image Processing 21(12), 4695-4708 (2012)


\bibitem{nadeem2019fuzzy} Nadeem, M., Hussain, A., Munir, A.: Fuzzy logic based computational model for speckle noise removal in ultrasound
images. Multimedia Tools and Applications pp. 1-18 (2019)


\bibitem{prasath2014image} Prasath, V.S., Delhibabu, R.: Image restoration with fuzzy coefficient driven anisotropic diffusion. In: International
Conference on Swarm, Evolutionary, and Memetic Computing, pp. 145-155. Springer (2014)


\bibitem{ratner2007image} Ratner, V., Zeevi, Y.Y.: Image enhancement using elastic manifolds. In: Image Analysis and Processing, 2007. ICIAP
2007. 14th International Conference on, pp. 769-774. IEEE (2007)


\bibitem{rudin2003multiplicative} Rudin, L., Lions, P.L., Osher, S.: Multiplicative denoising and deblurring: Theory and algorithms. In: Geometric Level
Set Methods in Imaging, Vision, and Graphics, pp. 103-119. Springer (2003)


\bibitem{shan2019multiplicative} Shan, X., Sun, J., Guo, Z.: Multiplicative noise removal based on the smooth diffusion equation. Journal of Mathe-
matical Imaging and Vision pp. 1-17 (2019)


\bibitem{dataset_eoportal} eoPortal: Sharing Earth Observation Resources: Kompsat-5. \url{https://directory.eoportal.org/web/eoportal/
satellite-missions/k/kompsat-5}


\bibitem{song2003fuzzy} Song, J., Tizhoosh, H.: Fuzzy anisotropic diffusion: a rule-based approach. In: Proceeding of the 7th World Multicon-
ference on Systemics, Cyebernetics and Informatics, pp. 241-246 (2003)

\bibitem{sun2016class} Sun, J., Yang, J., Sun, L.: A class of hyperbolic-parabolic coupled systems applied to image restoration. Boundary
Value Problems 2016(1), 187 (2016)

\bibitem{szmidt2000distances} Szmidt, E., Kacprzyk, J.: Distances between intuitionistic fuzzy sets. Fuzzy sets and systems 114(3), 505-518 (2000)

\bibitem{wang2004image} Wang, Z., Bovik, A.C., Sheikh, H.R., Simoncelli, E.P.: Image quality assessment: from error visibility to structural
similarity. Image Processing, IEEE Transactions on 13(4), 600-612 (2004)

\bibitem{yu2002speckle} Yu, Y., Acton, S.T.: Speckle reducing anisotropic diffusion. IEEE Transactions on image processing 11(11), 1260-1270
(2002)

\bibitem{zhou2015doubly} Zhou, Z., Guo, Z., Dong, G., Sun, J., Zhang, D., Wu, B.: A doubly degenerate diffusion model based on the gray level
indicator for multiplicative noise removal. IEEE Transactions on Image Processing 24(1), 249-260 (2015)

\bibitem{zhou2018nonlinear} Zhou, Z., Guo, Z., Zhang, D., Wu, B.: A nonlinear diffusion equation-based model for ultrasound speckle noise removal.
Journal of Nonlinear Science 28(2), 443-470 (2018)

\end{thebibliography}

\end{document}